\documentclass[aps,pra,10pt,twocolumn,floatfix,nofootinbib,superscriptaddress,longbibliography,nobibnotes]{revtex4-2}

\usepackage{amsmath}
\usepackage{amssymb}
\usepackage{mathtools}
\usepackage{wasysym}

\usepackage[T1]{fontenc}
\usepackage{amsfonts}
\usepackage{newtxtext}
\usepackage[varvw]{newtxmath}
\usepackage{dsfont}                 
\usepackage{bbold}                  
\usepackage[normalem]{ulem}         

\usepackage{enumerate}
\usepackage[shortlabels]{enumitem}

\usepackage[dvipsnames,table]{xcolor}
\usepackage{graphicx}
\graphicspath{{figs/}}              

\usepackage{physics}                
\usepackage{csquotes}               
\usepackage{comment}

\usepackage[caption=false]{subfig}
\usepackage[colorlinks=true,citecolor=cyan,linkcolor=magenta,filecolor=magenta]{hyperref}
\usepackage[capitalize]{cleveref}

\usepackage{orcidlink}

        \definecolor{AAcolor}{rgb}{0.7,0.1,0.4}

 \definecolor{TB}{rgb}{0,0.52,0.42}

\newcommand{\RNum}[1]{\uppercase\expandafter{\romannumeral #1\relax}}

\newcommand{\ordprod}{%
  \mathop{\overline{\prod}}\limits
}

\definecolor{ZG}{rgb}{1,0.2,0.2}

\begin{document}
\title{Delicate Wannier insulators}
\author{Zolt\'{a}n Guba\,\orcidlink{0000-0002-6130-1064}
}
\email{zoltan.guba@physik.uzh.ch}
\affiliation{Department of Physics, University of Zurich, Winterthurerstrasse 190, 8057 Zurich, Switzerland}
\author{Aris Alexandradinata\,\orcidlink{0000-0003-2235-497X}}    \email{aalexan6@ucsc.edu}
\affiliation{Physics Department and Santa Cruz Materials Center, University of California Santa Cruz, Santa Cruz, CA 95064, USA}
\author{Tom\'{a}\v{s} Bzdu\v{s}ek\,\orcidlink{0000-0001-6904-5264}}
\email{tomas.bzdusek@uzh.ch}
\affiliation{Department of Physics, University of Zurich, Winterthurerstrasse 190, 8057 Zurich, Switzerland}

\begin{abstract}
The defining feature of topological insulators is that their valence states are not continuously deformable to a suitably defined atomic limit without breaking the symmetry or closing the energy gap. When the atomic limit is given by symmetric exponentially-localized Wannier orbitals, one finds stable and fragile topological insulators characterized by robust bulk-boundary correspondence. More recently, delicate topological insulators (DIs) have been introduced, whose metallic states are guaranteed only at sharply terminated edges and surfaces. Although Wannierizable, their Wannier orbitals necessarily span multiple unit cells, thus refining the notion of the atomic limit. In this work, we extend delicate topological invariants from Bloch states to hybrid Wannier functions. The resulting models, dubbed delicate Wannier insulators (DWIs), are deformable to unicellular atomic limit in the absence of edges and surfaces; nevertheless, they exhibit obstructions to such deformations as well as topological boundary states in the presence of sharply terminated hinges and corners. We present a layering construction that allows us to elevate a DI in $d$ dimensions into a DWI in $(d\,{+}\,1)$ dimensions. We illustrate the phenomenology of DWIs by deploying the layering construction on three concrete models.
\end{abstract}

\maketitle

\section{Introduction}

Topological insulators (TIs) have garnered significant attention in condensed matter physics due to their unique electronic properties, characterized by insulating bulk and conductive edge channels robust against symmetry-preserving perturbations~\cite{Qi:2011,Hasan:2010}. 
Their unusual spectral and transport properties also motivate the search for a complete characterization of TIs.
Over the past two decades, various definitions of topologically nontrivial bands have been proposed, which in combination with the broad range of possible symmetry settings have resulted in extensive classification schemes.
Generally, these approaches rely on specifying so-called atomic limits, i.e., systems made of localized orbitals that are decoupled from each other, and then asking whether a given insulating Hamiltonian can be continuously deformed into an atomic limit without closing the energy gap and symmetry breaking.
Nevertheless, although tremendous progress has been achieved towards an exhaustive classification of topological insulators~\cite{Ryu:2010,Chiu:2016,Kruthoff:2017,Bradlyn:2017,Elcoro:2021,Brouwer}, the mathematical complexity of the problem remains so rich that novel types of topological insulators periodically emerge that break the formerly established categories.

Since many recently reported novel topological phases occur in special regimes (such as disordered crystals~\cite{Lappiere:2024}, amorphous lattices~\cite{Corbae:2023}, dissipative systems~\cite{Okuma:2023}, or strongly correlated Hamiltonians~\cite{Ohyama:2024}), one might be misled to deem the classification of TIs in the most elementary setting of crystalline lattices with negligible interactions to be fully solved.
However, this conclusion is wrong, as indeed illustrated by the sporadic discoveries of new classes of TIs, including higher-order~\cite{Schindler:2018}, fragile~\cite{Po:2018}, delicate~\cite{Nelson_prl}, multi-gap~\cite{Wu:2018,Lapierre_prr}, or boundary-obstructed~\cite{Khalaf} TIs. 
Indeed, each of these categories of TIs has provided an extension of topological band theory upon refining the formerly assumed conditions on crystal geometry, gap closing, or Hilbert space.
In this work, we extend these earlier developments by presenting a further category of topologically nontrivial insulators: \emph{delicate Wannier insulators} (DWIs).
We define DWIs as insulating Hamiltonians in which a delicate topological invariant is elevated from Bloch states to hybrid Wannier functions. 
We argue that these insulators are invisible to the formerly established classification techniques; nevertheless, certain DWIs exhibit higher-order boundary states that would appear mysterious (and considered accidental) without the formalization of their topological obstruction as expounded in the present work.

We find that the topological stability and bulk-boundary correspondence of DWIs exhibit certain similarities with higher-order TIs, fragile TIs, delicate TIs, and boundary-obstructed TIs; however, they also fundamentally differ from each of these categories of models.
To clarify these distinctive properties, let us reprise the key features of the just mentioned and formerly established classes of topological insulators.
First, higher-order TIs (HOTIs)~\cite{Schindler:2018,Schindler:2018b,Benalcazar:2017a,Benalcazar,Yang:2020} are TIs whose edges and surfaces are generically gapped and whose topological boundary states are restricted (potentially in the form of a filling anomaly~\cite{Benalcazar:2019}) to higher-order boundaries, i.e., hinges or corners (by `corner' and `hinge', we mean respectively a 0D boundary of a 2D system and a 1D boundary of a 3D system).
Many (though not all) investigated models of HOTIs are stable topological and visible to symmetry indicators (i.e., eigenvalues of occupied states at high-symmetry points~\cite{Po:2017}), and their bulk-boundary correspondence is robust against perturbations~\cite{Trifunovic:2019}.
While some models of DWIs which we present below exhibit topological modes on higher-order boundaries, their obstructions are not stable (they can be trivialized by enlarging the Hilbert space), they are not visible to symmetry indicators (their symmetry indicators are all trivial), and the topological modes can be removed (by relaxing the requirement of a sharp boundary termination).

From the perspective of the Hilbert space enlargement, three categories of TIs have been established. 
While stable TIs are robust against the addition of trivial bands to both the conduction (i.e.~unoccupied) and the valence (i.e.~occupied) sector, fragile TIs (FIs)~\cite{Alexandradinata,Bouhon:2019,Song} can only be trivialized by adding trivial bands to the valence (but not to the conduction) sector.
Both stable and fragile TIs lack decomposition into elementary band representations (EBRs)~\cite{Cano:2021}; consequently, they are detectable by symmetry indicators~\cite{Po:2017} and exhibit an obstruction to the construction of symmetric exponentially localized Wannier orbitals for spatial dimensions $d\geq 2$~\cite{Read:2017,Po:2018}. 
In contrast, delicate TIs (DIs)~\cite{Nelson_prb,Zhu:2023,Zhu:2023,Lim:2023,Kim:2024,Sun:2018,Park:2023,Bouhon:2024,Zhu:2024,Mo:2025} can be trivialized by adding a trivial band either to the conduction or to the valence subspace.
Although DIs are decomposable into EBRs and Wannierizable, their Wannier orbitals necessarily extend over multiple unit cells, leading to the refined notion of unicellular atomic limit~\cite{Nelson_prl}.
DIs are revealed by homotopy invariants of the Bloch Hamiltonian~\cite{Brouwer,Nelson_prl}, and they are observed~\cite{Chen_2024} (sometimes proved~\cite{Alexandradinata_prb,Nelson_prb}) to exhibit metallic states at sharply terminated edges and surfaces (by `edge' and `surface', we mean respectively a 1D boundary of a 2D system and a 2D boundary of a 3D system).
These states can be removed from the gap with a suitable symmetry-preserving boundary perturbation, i.e., when the boundary termination ceases to be sharp.
Some of the presently introduced DWIs exhibit conditionally robust metallic states at sharply terminated higher-order boundaries (corners and hinges); however, a distinction from DIs arises in that DWIs are trivial from the vantage point of homotopy theory. 
As a consequence, they are deformable to the unicellular atomic limit in the presence of periodic boundary conditions (PBC).
In addition, they cannot be trivialized by adding trivial bands to the conduction sector, reminiscent of FIs.

Finally, boundary-obstructed TIs (BOTIs)~\cite{Khalaf} correspond to Hamiltonians that can be continuously deformed (while preserving energy gap and symmetry) to atomic limit if adopting PBC but not in the presence of an open boundary termination.
The key idea is that while the bulk valence bands may be decomposable into EBRs of the bulk space group, the valence states near an open edge/surface may fail to be decomposable into EBRs of the edge/surface space group.
Alternatively, the decomposition into EBRs may exist, but it may not coincide with the underlying atomic positions (scenario referred to as an obstructed atomic limit~\cite{Bradlyn:2017,Cano:2022}; OAL).
This, in turn, enforces the appearance of metallic states at higher-order boundaries, i.e., hinges and corners~\cite{Benalcazar:2017a,Benalcazar}.
The DWIs are reminiscent of this phenomenology in that they are continuously deformable (while preserving energy gap and symmetry) to the unicellular atomic limit in the presence of PBC as well as in the presence of an open boundary in one direction, but not in the presence of open boundaries that meet at a higher-order boundary.
However, the first-order boundaries of DWIs are not characterized by nontrivial EBRs nor by OAL.

Our approach to introducing DWIs is based on elevating delicate topological invariants arising in formerly studied DIs from Bloch states (bulk energy eigenstates) to hybrid Wannier functions (bulk eigenstates of the projected position operator).
Therefore, to set the stage, we review in Sec.~\ref{sec:rtp} the paradigm model of a two-dimensional DI protected by mirror symmetry: the \emph{Returning Thouless Pump} (RTP) \emph{insulator}~\cite{Nelson_prl}, and we discuss its key phenomenology.
In particular, when ribbon geometry is adopted, the RTP insulator exhibits helical (i.e., two counterpropagating) metallic states at both edges, provided that a sharp boundary termination is ensured~\cite{Nelson_prb}.
Very recently, the theoretical concept of the RTP insulator has been experimentally realized in acoustic systems with synthetic dimensions~\cite{Cheng:2025,Mo:2025b}.

We continue in Sec.~\ref{sec:layered_rtp} by presenting our principal model of DWI: the \emph{layered RTP insulator} in 3D, and we inspect its spectrum under a broad range of boundary terminations. 
Similar to the RTP insulator in 2D, the layered RTP insulator is protected by mirror symmetry. 
In wire geometry with square cross-section, and assuming a sharp termination at the open boundaries, the layered RTP insulator exhibits helical modes at each of the four hinges, which are qualitatively similar to the helical edge states of the 2D RTP insulator in ribbon geometry.
We also introduce here our main tool for constructing a DWI in $(d+1)$ dimensions from any given DI in $d$ dimensions. 
Namely, we adopt a layering construction based on stacking alternating layers of insulators with pairwise canceling topological invariants. 
This method has been previously used by Ref.~\onlinecite{Khalaf} to construct models of BOTIs; for example, it generates the 2D Benalcazar-Bernevig-Hughes model with topological corner modes~\cite{Benalcazar:2017a} if applied to the Su-Schrieffer-Heeger model~\cite{Su:1980} which is the elementary model~of~OAL~in~1D.

To demonstrate the convenience of the layering construction, we use it to introduce two additional models of DWIs.
First, we discuss in Sec.~\ref{sec:layering1D} a simple model of a \emph{delicate topological 1D chain} that exhibits two in-gap states at each end.
Upon layering, one obtains a 2D DWI which in square geometry exhibits two in-gap states at each corner, i.e., a total of eight near-zero-energy topological modes.
This discussion is supplemented in Appendices~\ref{sec:1d-chain-two-bands} and~\ref{sec:1d-chain-three-bands} with a dedicated analysis of stable and delicate band topology of 1D chains symmetric under the operation of space inversion and time reversal.

As our final example, we review in Sec.~\ref{sec:layeringCDI} the recently introduced model of 2D DI dubbed \emph{Chern dartboard insulator} (labeled CDI$_2$)~\cite{Chen_2024}, whose valence band exhibits a quantized Chern number in individual quadrants of the Brillouin zone.
The authors of Ref.~\onlinecite{Chen_2024} reported that their model of $\textrm{CDI}_2$ in square geometry exhibits helical hinge modes on all edges (although we find that inclusion of additional terms in the bulk Hamiltonian, which preserve both the symmetry and the topology, may gap open an energy gap through hybridization of the edge modes). 
The corresponding \emph{layered CDI$_2$} is a DWI which (for the elementary choice of the bulk Hamiltonian) exhibits metallic helical hinge modes for two different orientations of the wire. 
We conclude the work in Sec.~\ref{sec:conclusion}, where we summarize our results and present several perspectives~for~future~works.

\section{Returning Thouless pump}
\label{sec:rtp}

\begin{figure}
\centering
\includegraphics[width=0.99\linewidth]{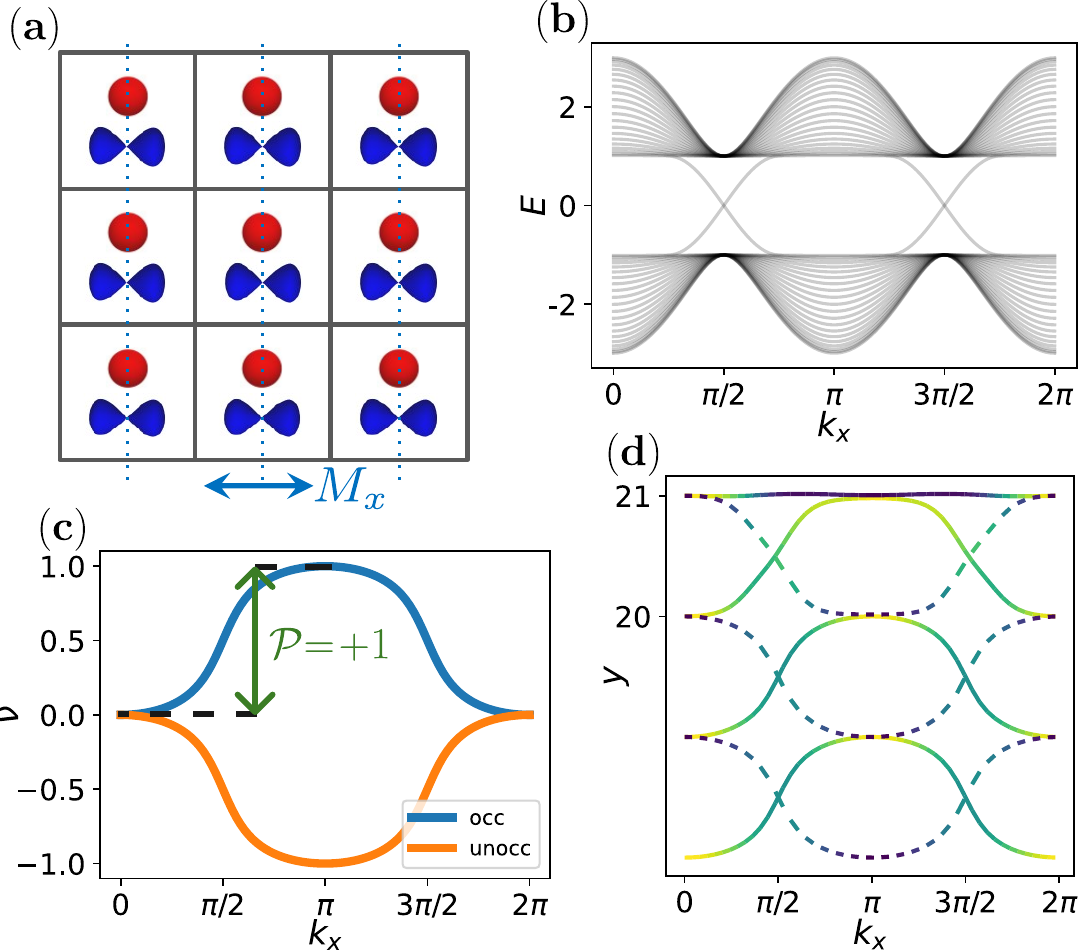}
\caption{The RTP model. 
(\textbf{a})~Real space lattice underlying the Hamiltonian in Eq.~(\ref{eq:rtp_model}). 
Each unit cell contains an $s$ and a $p$ orbital positioned along the axis of mirror symmetry $M_x$. 
(\textbf{b})~Band structure of the model in ribbon geometry with sharp open boundary condition in the $y$-direction. 
The crossing metallic bands are located at opposite edges of the system and carry a $\pi$-Berry phase. 
Data is obtained for $N_y = 21$ unit cells and $\alpha = 1$. 
The same parameters are used for the other panels as well. 
(\textbf{c})~The change of the bulk polarization from $k_x = 0$ to $\pi$ is an integer (in units of the lattice constant $a_y$).
(\textbf{d})~(adapted from Ref.~\onlinecite{Alexandradinata_prb}) Schematic figure of the spectrum of the projected position operator $\mathbb{P}y\mathbb{P}$ (solid) and $\mathbb{Q}y\mathbb{Q}$ (dashed). 
Here, $\mathbb{P}$ ($\mathbb{Q}=\mathbb{1}-\mathbb{P}$) is a projector to the occupied (unoccupied) subspace in the ribbon geometry at half-filling. 
The colors denote the orbital content of the eigenstates, with yellow (blue) implying a pure $s$-like ($p$-like) orbital character.
The topmost band of $\mathbb{P}y\mathbb{P}$ is anomalous as its character changes from $s$-like to $p$-like between $k_x = 0$ and $\pi$. 
}
\label{fig:basicRTP}
\end{figure}

We revisit the fundamental concept of the Returning Thouless Pump (RTP) in the context of energy bands. To streamline the discussion, we utilize a minimal two-band model that captures the essential features of the RTP invariant.
In this model, each unit cell contains two orbitals that carry opposite mirror eigenvalues, which we call the $s$ vs.~the $p$ orbital.
The orbitals are positioned along the same set of mirror-invariant lines on a 2D square lattice, Fig.~\ref{fig:basicRTP}(\textbf{a}). 
The Bloch Hamiltonian we consider is written as 
\begin{align}
\label{eq:rtp_model}
    \mathcal{H}_{\mathrm{RTP}}(\boldsymbol{k};\alpha) &= 
    \sin(2 k_x) \sigma_x 
    +  \sin(k_x) \sin(k_y) \sigma_y \nonumber \\
    &- (\cos(2 k_x) + \cos(k_y) + \alpha) \sigma_z,
\end{align}
where $\boldsymbol{k}=(k_x,k_y)$ is a two-dimensional momentum, and $\alpha$ is a parameter used to tune between the trivial and the topological phase of the model. 
The Hamiltonian obeys mirror symmetry 
\begin{equation}
\label{eq:rtp_mirrorsymmetry}
    M^{-1}_x \mathcal{H}_{\mathrm{RTP}}(k_x,k_y) M_x = \mathcal{H}_{\mathrm{RTP}}(-k_x,k_y) 
\end{equation}
where $M_x = \sigma_z$. 
The Hamiltonian in Eq.~(\ref{eq:rtp_model}) is constructed so that each half of the Brillouin zone ($k_x\,{\in}\,[0,\pi]$ vs.~$k_x\,{\in}\,[\pi,2\pi]$) 
contains one copy of the Qi-Wu-Zhang Chern-insulating model~\cite{Qi:2006}, but the opposite sign of the $\sin k_y \sigma_y$ term enforces cancellation of their Chern numbers.

Restricting our attention to $\alpha > 0$ and half-filling, the model is a semimetal for $\alpha_\textrm{c} = 2$ and an insulator otherwise. 
(The additional phases at $\alpha\leq 0$ are of no interest for the subsequent discussion.)  
The two insulating phases separated by the critical point at $\alpha = \alpha_\textrm{c}$ are equivalent from the perspective of symmetry indicators and both exhibit vanishing Zak-Berry phase~\cite{Zak:1989}.
Nevertheless, the insulating phase at $\alpha < 2$ is not deformable into an atomic limit with a $\boldsymbol{k}$-independent Bloch Hamiltonian without closing the energy gap (assuming we do not break the mirror symmetry and do not enlarge the dimension of the Hilbert space), hence it should be considered nontrivial.
This nontrivial aspect is revealed by a topological invariant formulated in Eq.~(\ref{eqn:RTP-invariant-def}).
In contrast, $\alpha >2 $ corresponds to a trivial insulator that can be deformed into such unicellular atomic limit. 
The distinct topological nature of the two gapped phases is reflected by the energy spectrum in the ribbon geometry. 
Imposing sharp open boundaries in the $y$-direction and periodic boundary condition in the $x$-direction, the topological phase exhibits metallic edge states that connect the valence band to the conduction band~\cite{Nelson_prb}, plotted in Fig.~\ref{fig:basicRTP}(\textbf{b}), whereas no such edge states arise in the trivial phase.

To characterize the topological phase, first note that mirror symmetry prevents the two basis orbitals from mixing along the high-symmetry lines $k_x=0$ and $k_x=\pi$, as $s$ and $p$ transform in different irreducible representations (irreps) of the little group. 
Second, the Hamiltonian is set up so that the lower-energy band is $s$-like along both $k_x\,{\in}\,\{0,\pi\}$ for both insulating phases. 
As a consequence, the polarization of the valence band in the $y$-direction matches at these $k_x$ values the location of the $s$ orbital,
\begin{equation}
\label{eq:rtp_equalpol}
    p^y(0) = y_s = p^y(\pi)~\mathrm{mod}~1.
\end{equation}
Here, we used that the polarization is defined only modulo lattice spacing, and we set the unit cell dimensions $a_x = 1 = a_y$ to unity. 
While Eq.~(\ref{eq:rtp_equalpol}) fixes the polarization at $k_x \,{\in}\, \{0,\pi\}$, it allows for a continuous \emph{change} of the polarization by an integer as $k_x$ is increased from $0$ to $\pi$.
The topological invariant
\begin{equation}
    \label{eqn:RTP-invariant-def}
    \mathcal{P} = \int_{0}^{\pi} d k_x \partial_{k_x} p^y(k_x)\in\mathbb{Z}
\end{equation}
allows us to distinguish the two insulating phases of the model in Eq.~(\ref{eq:rtp_model}).
Namely, as plotted in Fig.~\ref{fig:basicRTP}(\textbf{c}), in the case of $\alpha<2$ the polarization increases by $\mathcal{P}=+1$ between $k_x = 0$ and $k_x = \pi$, whereas $\mathcal{P}=0$ for $\alpha > 2$.
Note that the mirror symmetry implies that the change in polarization is reverted in the complementary half-BZ, implying a vanishing Chern number over the full BZ.
Owing to the returning nature of the polarization over the full BZ, the invariant $\mathcal{P}$ is called returning Thouless pump~\cite{Nelson_prb,Alexandradinata:2024} -- a name motivated by the promotion of $k_x$ into an adiabatic pump parameter~\cite{Thouless:1983}.

Returning Thouless pump (RTP) is quantized due to the validity of Eq.~(\ref{eq:rtp_equalpol}), which, in turn, is imposed by the opposite mirror symmetry of the conduction and valence band. 
The quantization is lost when the mirror symmetry is broken, meaning that RTP is an example of a symmetry-protected topology.
Less trivially, the quantization is also lost if a third band is added to the unoccupied sector whose mirror eigenvalue matches that of the occupied band (i.e., $s$) at either high-symmetry line. 
This is because under such conditions the mirror symmetry does not forbid the mixing of the two $s$ orbitals, so that the polarization at $k_x\,{\in}\,\{0,\pi\}$ can no longer be related to the position of a specific orbital; therefore, $\mathcal{P}$ can be continuously tuned to zero.
The same trivialization occurs when adding a band to the occupied sector whose mirror eigenvalue matches that of the unoccupied band (i.e., $p$). 
That the mirror eigenvalue of all occupied bands should be opposite to that of all the unoccupied bands for RTP to be a topological invariant has been dubbed the `mutually-disjoint condition' by Ref.~\onlinecite{Nelson_prb}.
That RTP can be trivialized by a suitable (symmetry-conditioned) enlargement of the Hilbert space is the hallmark of (symmetry-protected) delicate topology.

The existence of in-gap states in the ribbon geometry can be motivated by utilizing the projected $y$ operator in the ribbon geometry, $\mathbb{P}y\mathbb{P}$. 
Here, $\mathbb{P} = \mathbb{P}(k_x)$ is a momentum-resolved projector onto the occupied states in the ribbon geometry. 
To uniquely define the projector to the occupied subspace, one requires the existence of a global energy gap (i.e., including on the boundaries) in the spectrum around the Fermi energy. 
In the case of delicate phases, including the Hamiltonian in Eq.~(\ref{eq:rtp_model}), this can be achieved by a suitable boundary perturbation~\cite{Alexandradinata_prb}.
The bulk part of the $\mathbb{P}y\mathbb{P}$ spectrum and the $k_x$-resolved polarization $p^y(k_x)$ are in a one-to-one correspondence in a periodic system~\cite{King-Smith:1993,Marzari:1997}, but this relation breaks down near the edges of the ribbon.

The spectrum of the $\mathbb{P}y\mathbb{P}$ operator in the topological phase near the top edge of the ribbon is plotted with solid lines in Fig.~\ref{fig:basicRTP}(\textbf{d}).
While the bulk bands mimic the bulk polarization in Fig.~\ref{fig:basicRTP}(\textbf{c}), with the $\mathbb{P}y\mathbb{P}$ eigenvalue increasing by $\mathcal{P}=+1$ as $k_x$ increases from $0$ to $\pi$, the topmost band appears flat.
This anomalous aspect of the topmost $\mathbb{P}y\mathbb{P}$ band arises because it cannot mimic the returning pump profile as there are no additional layers physically present on top. 
Furthermore, since the $s$ orbitals near the top edge are all `used up' to constitute the $\mathbb{P}y\mathbb{P}$ bands at $k_x=\pi$ further from the edge, there is no $s$ orbital into which the topmost $\mathbb{P}y\mathbb{P}$ band can evolve as $k_x$ increases from $0$ to $\pi$.
Hence, the orbital character of the topmost band is forced to change from $s$-like to $p$-like. 
This results in an overall $\pi$ Berry phase for the states near the top edge, which can be extracted irrespective of the boundary termination using a Wannier cut~\cite{Nelson_prb,Trifunovic}. 
The existence of the in-gap state in the presence of a sharply terminated edge (i.e., in the absence of perturbations near the edge) follows from the fact that on-site energy of the $s$ ($p$) orbitals is negative (positive).\footnote{This argument is discussed in detail in Appendix~O of Ref.~\cite{Nelson_prb}.}

\section{Layered RTP insulator}
\label{sec:layered_rtp}

In this section, we turn to the main model investigated in this work: the layered RTP insulator. 
We construct the model in Sec.~\ref{sec:layering} by coupling layers of RTP insulators along a new spatial direction, following the approach introduced in Ref.~\onlinecite{Khalaf}. 
We also discuss here the spectral properties of the model, pointing out the appearance of metallic helical modes at sharply terminated hinges.
We next analyze in Sec.~\ref{sec:lay-RTP-topology} the symmetries of the layered RTP insulator and the topology of its Wannier bands. 
Since the topological invariant characterizing the Wannier bands of the layered RTP model is delicate, we describe it as a delicate Wannier insulator (DWI).
We further expose in Sec.~\ref{sec:hinge-Berry} that the helical hinge modes are associated with an anomalous $\pi$ Berry phase.
Finally, in Sec.~\ref{sec:layered-RTP-PD&obst} we explore the phase diagram of the model by identifying the energy gap closings in the bulk and on the surfaces. 
We observe that the surface gap closings of the layered RTP model occur simultaneously with the closings of the Wannier gap in the corresponding direction.
We also comment here on the boundary-obstructed aspect of the model.

\subsection{Layering construction}\label{sec:layering}

\begin{figure}[t]
    \centering
    \includegraphics[width=0.9\linewidth]{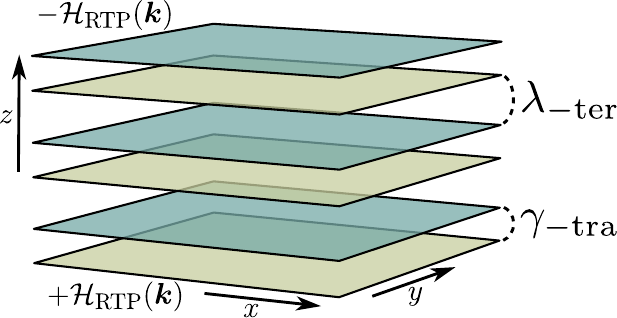}
    \caption{
    Construction of the layered RTP insulator [Eq.~(\ref{eqn:layered-RTP-Hamiltonian}), Sec.~\ref{sec:layering}]. 
    The dark resp.~light green planes denote copies of the RTP insulator with opposite signs. 
    Intra- and interlayer couplings are denoted with~$\gamma_\textrm{-tra}$~and~$\lambda_\textrm{-ter}$. 
    We adopt an equivalent layering construction to also obtain our other examples of DWIs, namely the layered delicate topological 1D chain [Eq.~(\ref{eqn:layered-chain-model}), Sec.~\ref{sec:deli-chain-layered}] and the layered Chern dartboard insulator [Eq.~(\ref{eqn:layered-CDI-Ham}), Sec.~\ref{sec:CDI-layered}].
    }    
    \label{fig:layeredRTP_construction}
\end{figure}

The layered RTP insulator is assembled as follows. 
First, we take two copies of the RTP insulator with opposite value of the invariant $\mathcal{P}$ in Eq.~(\ref{eqn:RTP-invariant-def}) and with the orbital content flipped between the conduction and the valence bands.
We achieve this by using $\mathcal{H}_\textrm{RTP}(\boldsymbol{k};\alpha)$ from Eq.~(\ref{eq:rtp_model}) as one copy, and the negative $-\mathcal{H}_\textrm{RTP}(\boldsymbol{k},\alpha)$ as the other copy.
The sum of these two models exhibits no net change of polarization over half-BZ; in fact, the total value of $\mathcal{P}$ is, strictly speaking, not well-defined due to the loss of the mutually-disjoint condition~\cite{Nelson_prb}.  
Then, as illustrated in Fig.~\ref{fig:layeredRTP_construction}, we couple many such pairs of RTP insulators with intra- and interlayer couplings $\gamma_\textrm{-tra}$ resp.~$\lambda_\textrm{-ter}$ in the $z$-direction, forming a three-dimensional slab. 
The resulting Bloch-Hamiltonian is written as\footnote{
When constructing Eq.~(\ref{eqn:layered-RTP-Hamiltonian}), we have adopted the `periodic' Bloch convention, for which the Bloch Hamiltonian $\mathcal{H}(\boldsymbol{k})$ is $\boldsymbol{k}$-periodic in reciprocal lattice vectors. 
This means we are implicitly assigning both layers within a unit cell the same $z$ coordinate. Inclusion of the vertical separation between the layers alters the specific form of the Hamiltonian but does not qualitatively change any of the presented findings. 
}
\begin{align}
\label{eqn:layered-RTP-Hamiltonian}
    \mathcal{H}_{\,\textrm{RTP}}^{\mathrm{layered}}(\boldsymbol{k}) \!=\! \tau_z \otimes \mathcal{H}_{\mathrm{RTP}}(\boldsymbol{k};\alpha)  + \lambda_\textrm{-ter} \sin k_z \, \tau_y  \otimes  \sigma_0  + \nonumber \\
    + (\gamma_\textrm{-tra} + \lambda_\textrm{-ter} \cos k_z) \, \tau_x  \otimes  \sigma_0 , 
\end{align}
where the $\tau_i$ operators act on the sublayer degree of freedom, while $\sigma_i$ are the orbital Pauli matrices inherited from Eq.~(\ref{eq:rtp_model}).

To gain intuition about the phase diagram of the layered RTP model in Eq.~(\ref{eqn:layered-RTP-Hamiltonian}), let us first investigate two particular limits.
On one hand, when $\lambda_\textrm{-ter}=0$ and $\gamma_\textrm{-tra} \neq 0$, the system consists of decoupled layers of pairs of RTP insulators carrying opposite value of $\mathcal{P}$. 
Such a system corresponds to a trivial phase irrespective of the invariant $\mathcal{P}$ associated with the constituent RTPs. On the other hand, when $\gamma_\textrm{-tra} = 0$ and $\lambda_\textrm{-ter} \neq 0$, the system exhibits a decoupled RTP-insulator layer both at the bottom and at the top of the slab. 
Assuming the single RTP-insulator layer, analyzed in Sec.~\ref{sec:rtp}, is in the topologically nontrivial phase, its sharply-terminated edges host metallic modes. 
Within the layered geometry, sharply-terminated edges of the outermost layers translate into sharply-terminated hinges, equipping the metallic boundary modes with a higher-order topological character.
We anticipate the metallic hinge modes to persist in the spectrum until either the bulk or a surface of the model exhibits an energy gap closing.

To demonstrate the persistence of the metallic hinge modes beyond the decoupled limit, let us further fix  $\gamma_\textrm{-tra}= 0.5$ and $\lambda_\textrm{-ter} = 1$, with $\alpha = 1$ ensuring nontrivial topology of the constituent RTP layers.
The energy bands and the Wannier bands of the model for this choice of parameters are shown in Fig.~\ref{fig:layeredRTP_bands}. 
We show that the model is insulating in the bulk and on the surfaces at half-filling. 
To do this, we first diagonalize the Hamiltonian with PBCs in all directions; then, we create open surfaces in the $y$- and separately in the $z$-directions and we diagonalize the Hamiltonian in these settings. 
The energy levels in each of these three geometries exhibit a gap separating the valence states from the conduction states (not plotted).

Finally, we enforce open boundary conditions (OBCs) in both the $y$ and the $z$-direction, thus forming a wire in which $k_x$ remains a good quantum number. 
The spectrum of the resulting wire, plotted in 
Fig.~\ref{fig:layeredRTP_bands}(\textbf{a}),
exhibits two doubly-degenerate metallic bands inside the bulk and surface energy gap, which we find to be localized at the hinges of the wire. 
The double degeneracy follows from $\pi$-rotation symmetry about the $x$-axis, which relates opposite hinges of the wire and is represented by $C_{2x}=\tau_x \otimes \sigma_y$ in the bulk.
These hinge-localized states belong to the top and bottom RTP layers and, as such, are expected to inherit the anomalous $\pi$-Berry phase.

\subsection{Topology of Wannier bands}\label{sec:lay-RTP-topology}

\begin{figure}
\centering
\includegraphics[width=1\linewidth]{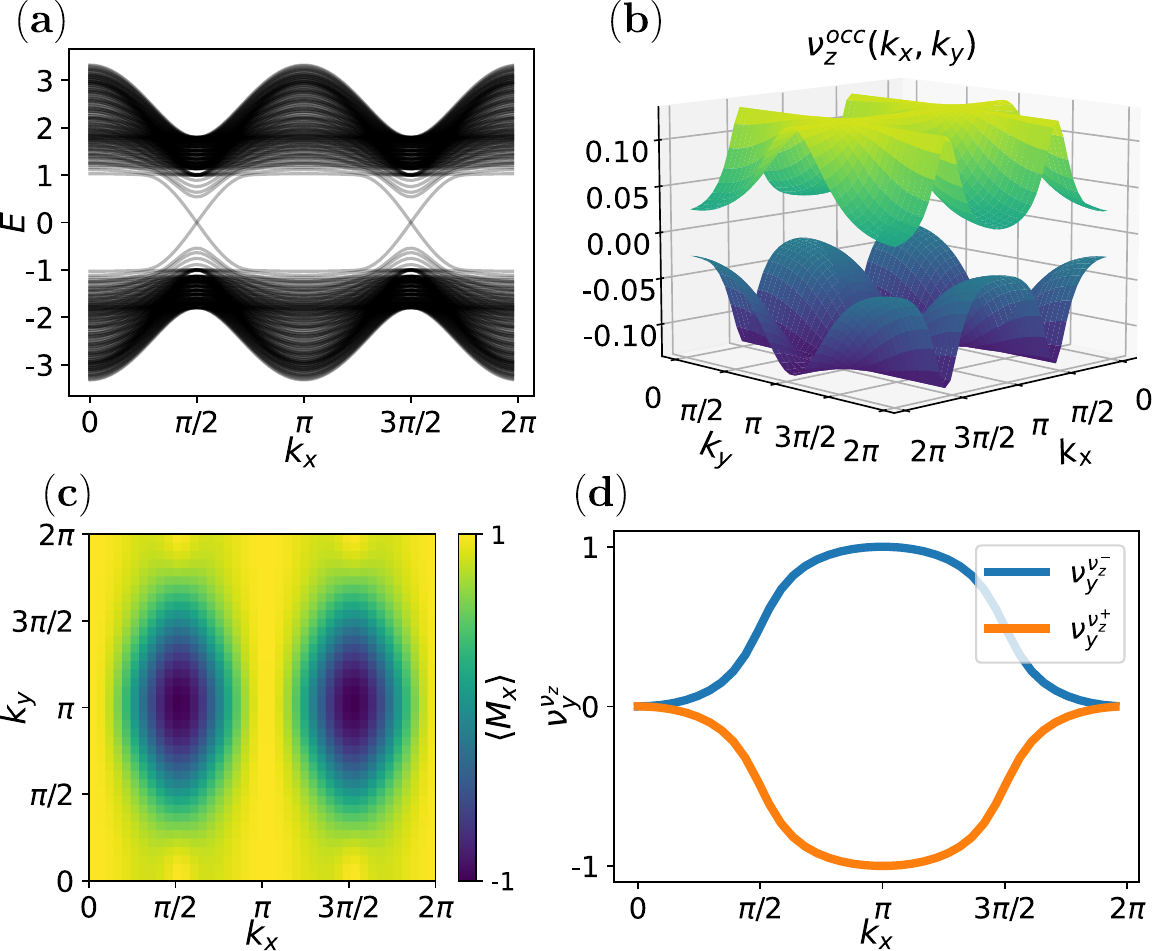}
\caption{
Energy bands and Wannier bands of the layered RTP model in Eq.~(\ref{eqn:layered-RTP-Hamiltonian}) under various boundary conditions for $\alpha = 1, \gamma_\textrm{-tra} = 0.5$, and $\lambda_\textrm{-ter} = 1$.
(\textbf{a})~Energy spectrum in the wire geometry with $N_y = 12 = N_z$ unit cells and with OBC in both the $y$ and the $z$-direction. We observe the appearance of helical hinge modes inside the gap. 
(\textbf{b})~Wannier bands of the occupied subspace computed in the $z$-direction. 
These Wannier bands mimic the topology of the energy bands of the RTP model. 
Data is generated for $N_x = N_y = N_z = 41$.
(\textbf{c})~The expectation value of $M_x$ for the lower Wannier band $\nu^-_{z}$. The orbital content of $\nu^-_z$ corresponds to that of the occupied energy band of 
$\mathcal{H}_{\mathrm{RTP}}$.
(\textbf{d})~Polarization of the Wannier bands. 
The lower (upper) Wannier band $\nu^{-}_z$ ($\nu^{+}_z$) has RTP invariant $\mathcal{P} = +1$ ($-1$) inherited from the topology of the constituent RTP layers.
} 
\label{fig:layeredRTP_bands}
\end{figure}

We next turn our attention from energy bands to Wannier bands. 
We compute the charge center of hybrid Wannier functions localized in the $z$-direction, which is achieved by parallel transport of the occupied Bloch states along noncontractible loops in the stacking ($k_z$-)direction. 
The resulting Wannier bands $\nu_z^\textrm{occ}(k_x,k_y)$, plotted in  
Fig.~\ref{fig:layeredRTP_bands}(\textbf{b}), are separated by a spectral gap at both $\nu_z = 0$ and $\nu_z=\tfrac{1}{2}$, in units where the value `$1$' corresponds to the lattice period in the $z$ direction.

Importantly, we find that the pair of Wannier bands exhibit the topology of the constituent RTP layers with a non-vanishing value of the RTP invariant. 
Specifically, while the nontrivial topology of the RTP insulator is revealed by the quantized change of polarization of the (either occupied or unoccupied) energy bands [Fig.~\ref{fig:basicRTP}(\textbf{c})], the nontrivial topology of the layered RTP insulator is revealed by a quantized change of the \emph{nested} Wilson loop~\cite{Benalcazar:2017a} of the (either lower or upper) Wannier band [Fig.~\ref{fig:layeredRTP_bands}(\textbf{d})].
Since this is a delicate topological invariant, we describe the layered RTP model as a \emph{delicate Wannier insulator} (DWI).
Owing to the heuristic correspondence between the Wannier bands and the surface spectra of semi-infinite systems~\cite{Fidkowski:2011,Benalcazar}, the obtained Wannier bands tentatively serve as a proxy for the occupied subspace of the top and the bottom surface in the slab geometry.

The nontrivial topology of the layered RTP model is protected by the mirror symmetry $M_x$ inherited from the constituent copies of the RTP insulators. 
The symmetry acts as 
\begin{equation}
\label{eq:layeredrtp_mirrorsymmetry}
    M^{-1}_x \mathcal{H}_{\,\textrm{RTP}}^{\mathrm{layered}}(\boldsymbol{k}) M_x = \mathcal{H}_{\,\textrm{RTP}}^{\mathrm{layered}}(M_x\boldsymbol{k}),
\end{equation}
with the mirror-symmetry operator $M_x = \tau_0 \otimes \sigma_z $ and with $M_x \boldsymbol{k} = (-k_x,k_y,k_z)$.  
Due to the layer construction with the alternating signs of $\mathcal{H}_\textrm{RTP}(\boldsymbol{k},\alpha)$, the occupied subspace contains both a mirror-even ($s$) orbital and a mirror-odd ($p$) orbital.
The Wannier bands are gapped and satisfy the mutually-disjoint condition.
Namely, the lower Wannier band is $s$-like, while the upper one is $p$-like at both high-symmetry lines $k_x=0$ and $k_x=\pi$. 
This leads to a quantized change in the polarization of either Wannier band between $k_x=0$ and $\pi$, due to the same argument as used to explain the quantization of the RTP invariant $\mathcal{P}$ in Eq.~(\ref{eqn:RTP-invariant-def}).
The topology of the Wannier band can be equivalently intepreted as the Chern number (or the winding number) of the lower Wannier band $\nu_z^-$ accumulated over the half-BZ $[0,\pi]$. 
A nontrivial value of this winding number implies that the states $\ket{\nu^z_-(k_x,k_y)}$ constituting the lower Wannier band enclose the Bloch sphere as the momentum $(k_x,k_y)$ is tuned over half the Brillouin zone, which is reflected in the pattern of the expectation value 
$\bra{\nu_z^-}M_x\ket{\nu_z^-}$ for all $(k_x,k_y)$ in Fig.~\ref{fig:layeredRTP_bands}(\textbf{c}).

Interestingly, the Wannier bands exhibit a non-trivial RTP invariant also for $\gamma_\textrm{-tra}/\lambda_\textrm{-ter}>1$ (while keeping $\alpha<2$) when \emph{no} hinge modes are present in the specified wire geometry. 
We clarify in Sec.~\ref{sec:layered-RTP-PD&obst} that, in this regime, the model can still be regarded as a DWI; however, the appearance of the helical hinge modes is conditioned by adopting a different choice of the $z$-boundary termination.

\subsection{Hinge Berry phase}\label{sec:hinge-Berry}

\begin{figure}
    \centering
    \includegraphics[width=0.9\linewidth]{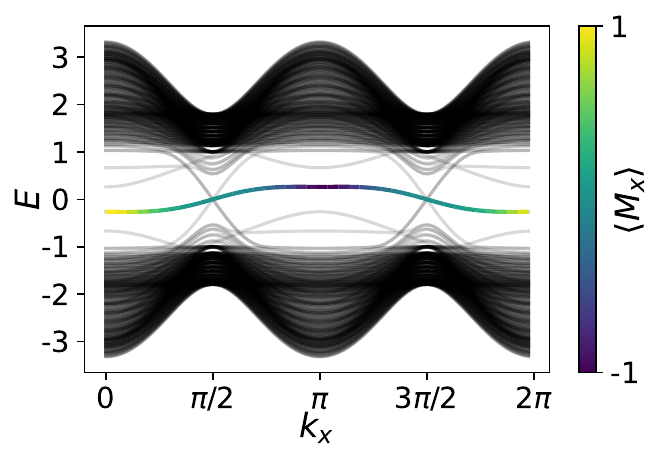}
    \caption{
    Revealing the topological nature of the hinge-localized states of the layered RTP model. 
    A perturbation localized to a hinge results in detachment of the metallic hinge mode from the bulk and surface states.
    The detached band is colored according to the expectation value of the mirror operator $M_x$. 
    The orbital character of the band changes from $s$-like to $p$-like, implying that it carries a $\pi$ Berry phase.
    Data is obtained for $\alpha = 1$, $\gamma_\textrm{-tra} = 0.5$, $\lambda_\textrm{-ter} = 1$, $N_y = 12 = N_z$.
    }
    \label{fig:layeredRTP_detachedwire}
\end{figure}

Due to the delicate nature of the topology of the layered RTP model, the hinge-localized modes interpolating between the conduction and the valence bands are not robust if the hinge termination is not sharp. 
We exemplify this by detaching one hinge mode from the bulk states using a suitable local perturbation. 
The specific perturbation used in our calculations corresponds to rescaling the selected elements of the Hamiltonian matrix $H(k_x)$ in the wire geometry as
\begin{equation}
    H_{ij} \mapsto \mu \cdot H_{ij},\label{eqn:hinge-perturbation}
\end{equation}
where $i$ and $j$ are degrees of freedom located inside the unit cells adjacent to a hinge (i.e., four orbitals in each layer), and $0 < \mu < 1$ is a number, which we set to $\mu=0.2$.  
The resulting energy spectrum is shown in Fig.~\ref{fig:layeredRTP_detachedwire}.

The spectral detachment of the hinge mode not only manifests the expected delicate nature of the studied topology, but it also allows us to determine the mirror eigenvalue of the hinge mode at high-symmetry momenta $k_x \,{\in}\,\{0,\pi\}$. 
This, in turn, enables us to inspect the orbital character of the detached band. 
In Fig.~\ref{fig:layeredRTP_detachedwire}, the detached hinge mode is colored according to its orbital character (i.e., the expectation value of $M_x$) for all momenta in the wire Brillouin zone.  
The change in the orbital character between $k_x = 0$ and $\pi$ implies a Berry phase $\phi_\textrm{B}^\textrm{hinge}=\pi$ associated with the hinge. 
This quantized Berry phase of the layered RTP model is inherited from the quantized Berry phase associated with the edges of the constituent~RTP~layers.\footnote{
The presented determination of the quantized hinge Berry phase is not biased by the choice of perturbation applied near the hinge. 
We elaborate on this issue in the concluding remarks in Sec.~\ref{sec:conclusion}.
}
In the following text, we consider the presence of the helical hinge modes as an indication of the anomalous hinge Berry phase.

\subsection{Phase diagram and boundary obstruction}\label{sec:layered-RTP-PD&obst}

\begin{figure}
    \centering
    \includegraphics[width=0.99\linewidth]{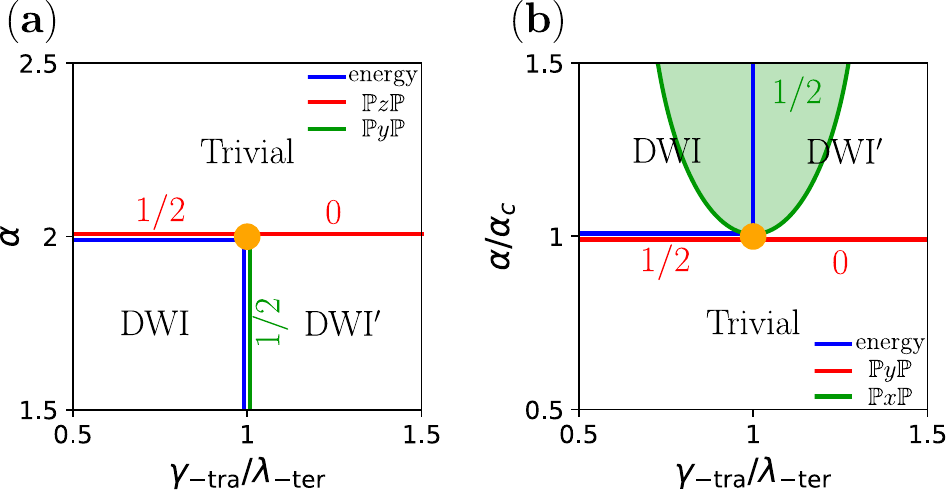}
    \caption{
    (\textbf{a})~Phase diagram of the layered RTP model in Eq.~(\ref{eqn:layered-RTP-Hamiltonian}) [Sec.~\ref{sec:layered_rtp}] and (\textbf{b}) phase diagram of the layered 1D chain in Eq.~(\ref{eqn:layered-chain-model}) [Sec.~\ref{sec:layering1D}].
    The blue lines correspond to closings of the energy gap at first-order boundaries [i.e., surfaces in (\textbf{a}) and edges in (\textbf{b})]
    in the considered geometry.
    Red and green lines (and the green region in the right panel) indicate closing of the Wannier gaps in the direction specified by the color legend (insets) and at the eigenvalue $\nu_j=0$ or $\nu_j=1/2$ as specified by the inscription next to the transition lines; $\mathbb{P}$ in the legend is projector onto occupied states. 
    Orange dots indicate a bulk energy gap closing.
    The regions labeled as DWI and DWI' are both topological in the sense that they exhibit a delicate topological invariant of the Wannier bands; however, only the region labeled DWI is topological in the sense of exhibiting higher-order boundary states in the specified wire resp.~square geometry (see the discussion in the last paragraph of Sec.~\ref{sec:layered-RTP-PD&obst}).
    The region labeled~`triv.'~is~trivial.
    }
    \label{fig:layeredRTP_phasediag}
\end{figure}

Let us address the phase diagram of the layered RTP model in Eq.~(\ref{eqn:layered-RTP-Hamiltonian}).
The phase diagram is displayed in Fig.~\ref{fig:layeredRTP_phasediag}(\textbf{a}) as a function of $\alpha$ and $\gamma_\textrm{-tra} / \lambda_\textrm{-ter}$, where
$\alpha$ is the parameter of the constituent RTP-insulator layers and
$\gamma_\textrm{-tra} / \lambda_\textrm{-ter}$ is the ratio of the intra- and interlayer couplings. 
The diagram indicates three phases.
First, the region $\alpha < 2$ and $\gamma_\textrm{-tra} / \lambda_\textrm{-ter} < 1$ is a DWI whose Wannier bands exhibit an RTP invariant and which exhibits helical modes at sharply terminated hinges in the specified wire geometry.
The DWI region of the phase diagram is demarcated by blue lines, which indicate closing of the energy gap at one of the surfaces adjacent to the hinge; the two lines meet at a point (orange dot) that corresponds to closing of the bulk energy gap.
Interestingly, the lines of closing the surface energy gap coincide with closing of the Wannier gap at $\nu=1/2$ in the direction normal to the corresponding surface [red and green lines in Fig.~\ref{fig:layeredRTP_phasediag}(\textbf{a})]. 
Next, the region with $\alpha >2$ (irrespective of the choice of $\gamma_\textrm{-ter}/\lambda_\textrm{-tra})$ is trivial; here, the Wannier bands carry zero RTP invariant and no helical modes are present at the wire hinges. 
Finally, the region with $\alpha<2$ and $\gamma_\textrm{-tra}/\lambda_\textrm{-ter}>1$ is marked as DWI$'$. 
In this region, which is characterized by a swapped dimerization pattern of the alternating layer, no helical modes arise at sharply terminated hinges in the specified wire geometry, even though the Wannier bands exhibit~a~nontrivial~RTP~invariant.

We next inspect the nature of the topological obstruction in the DWI phase of the layered RTP model.
Note that the bulk occupied subspace is decomposable into two line bundles (i.e., two unit-rank projectors over the 3D Brillouin zone) which are trivially permuted by the mirror symmetry.
An explicit choice of such a decomposition is provided by the eigenstates constituting the gapped Wannier bands $\nu_z^\textrm{occ}(k_x,k_y)$ [Fig.~\ref{fig:layeredRTP_bands}(\textbf{b})] and by their parallel transport~\cite{Soluyanov:2012} in the $k_z$-direction.
In addition, each of the two line bundles has a trivial Chern class (over the 3D Brillouin zone); therefore, the crystallographic splitting theorem guarantees that the occupied subspace is a band representation~\cite{Alexandradinata}.
It follows that the occupied bands are trivial from the vantage point of $K$-theory~\cite{Kruthoff:2017}, equivariant classification of vector bundles~\cite{Alexandradinata}, and symmetry indicators~\cite{Po:2017,Bradlyn:2017}.
Nonetheless, in the presence of a sharply terminated hinge, i.e., a higher-order boundary, metallic modes connecting valence bands to conduction bands appear. 
As revealed in Fig.~\ref{fig:layeredRTP_detachedwire}, the orbital character of the hinge modes is \emph{mixed}.
Such orbital character is compatible with a representation in terms of symmetric localized orbitals placed at the unit cell \emph{boundary}, if the unit cell is chosen to have the basis $s$ and $p$ orbitals at the center. This exemplifies the emergence of an OAL~on~the~hinge.

\begin{figure}[t]
    \centering
    \includegraphics[width=0.99\linewidth]{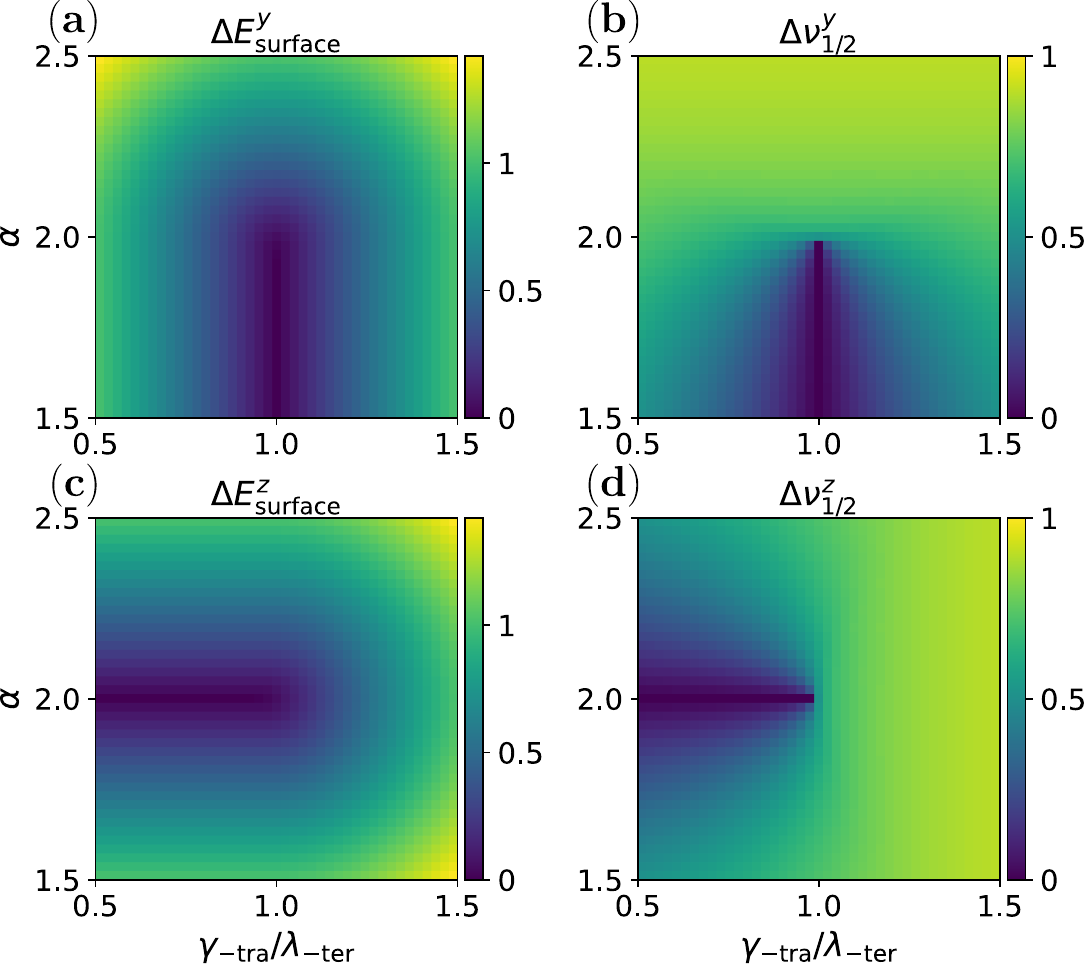}
    \caption{
    Spectral gaps of the layered RTP model [Eq.~(\ref{eqn:layered-RTP-Hamiltonian})] in slab geometry. 
    (\textbf{a})~The surface energy gap $\Delta E^y_\textrm{surface}$ of a slab with sharp open boundaries in the $y$-direction. 
    The gap closes at $\gamma_\textrm{-tra}/\lambda_\textrm{-ter}=1$ for $\abs{\alpha}<2$.
    (\textbf{b}) The observed closing of the energy gap coincides with the closing of the gap $\Delta \nu_{1/2}^y$ of the Wannier bands at $\nu = \tfrac{1}{2}$ in the $y$-direction.
    Panels (\textbf{c},\textbf{d}) display the analogous data for the $z$-direction. 
    Here, the gaps $\Delta E^z_\textrm{surface}$ and $\Delta \nu^z_{1/2}$ close at $\alpha = 2$ for $\abs{\gamma_{\textrm{-tra}}/\lambda_{\textrm{-ter}}}<1$.
    We refer to the simulteneity of the gap closings in the pairs of spectra as the concurrent-gap condition.
    Observe also that for a specified slab geometry, whether in panel (\textbf{a}) or (\textbf{c}), any two choices of parameters that exhibit a gapped energy spectrum can be connected without closing the energy gap in the slab.
    The data is obtained for slabs with $N_z = 52$ layers and with a grid of $N_x = N_y = 52$ momenta in the surface Brillouin~zone.
    }
    \label{fig:layered-RTP-gaps}
\end{figure}

We further address the question why the appearance of metallic states at sharply terminated hinges is bounded by the Wannier gap closings for the projected-$z$ and projected-$y$ operators. 
This holds true if the model satisfies a \textit{concurrent-gap condition}, meaning a concurrent (i.e., simultaneous) closure of the surface energy gap and of the Wannier gap in the same direction.
In more detail, we observe that the layered RTP model exhibits closing of the energy gap with sharply terminated surfaces in the $j$-direction ($j\in \{y,z\}$) simultaneously with the closing of the Wannier gap at $\nu = \tfrac{1}{2}$ for the projected position operator in the $j$ direction.
To illustrate this correspondence, we plot in Fig.~\ref{fig:layered-RTP-gaps} the surface energy gaps [panels (\textbf{a},\textbf{c})] and the Wannier gaps at $\nu=\tfrac{1}{2}$ [panels (\textbf{b},\textbf{d})] over the range of parameters $\alpha \in [1.5,2.5]$ and $\gamma_\textrm{-tra}/\lambda_\textrm{-ter} \in [1.5,2.5]$ of the phase diagram in Fig.~\ref{fig:layeredRTP_phasediag}(\textbf{a}).
The simultaneity of the specified gap closings has also been observed and utilized in the context of a certain class of HOTIs~\cite{Benalcazar}.
While it was formerly proved that the spectral \emph{flow} in the open boundary spectrum of a crystalline insulator is deformable to the spectral flow of the Wannier spectrum in the corresponding direction~\cite{Fidkowski:2011}, thus allowing to explain the bulk-boundary correspondence in a class of first-order TIs, the concurrent-gap condition is a \emph{stronger} statement that has not been mathematically established under suitably specified conditions.
While the presented layered RTP model satisfies (within human-eye precision) the concurrent-gap condition, it was shown in Ref.~\cite{Yang:2020} that this condition breaks down for generic models, such as upon the inclusion of next-to-nearest neighbor hoppings.

Putting aside the issue of genericity, we explore a further implication of assuming a concurrent-gap condition for the layered RTP insulator in Eq.~(\ref{eqn:layered-RTP-Hamiltonian}). 
Namely, in the presence of two sharply terminated open surfaces that meet at a hinge, it is not possible to continuously deform the phase characterized by the helical hinge modes (i.e., DWI) to a phase devoid of the helical hinge modes (either DWI$'$ or trivial) without closing the energy gap at one of the open surfaces. 
This phenomenology implies a boundary obstruction reminiscent of BOTIs~\cite{Khalaf}.
Specifically, the energy gap could close either on the $z$-terminated surface (which corresponds to trivialization of the RTP invariant within the individual constituent layers by tuning $\alpha$) or on the $y$-terminated surface (which corresponds to changing the dimerization pattern of the constituent layers by tuning $\gamma_\textrm{-ter}/\lambda_\textrm{-tra}$). 
In contrast, in slab geometry (whether finite in the $y$- or in the $z$-direction), one of the blue transition lines in Fig.~\ref{fig:layeredRTP_phasediag}(\textbf{a}) becomes absent.
This allows us to relate any two points in the phase diagram without encountering a closing of the energy gap anywhere in the system, as indeed confirmed by the energy gap data in Fig.~\ref{fig:layered-RTP-gaps}(\textbf{a}) resp.~(\textbf{c}).
In particular, we can reach the limit where the constituent layers are fully decoupled and their RTP invariant is trivial, in which case the model is deformable to the unicellular atomic limit~\cite{Nelson_prb}.
Therefore, with the specified constraints, we find that the layered RTP model is deformable to the unicellular atomic limit without closing of the energy gap in the slab geometry, but not in the wire geometry.

When additional terms are included in the layered RTP insulator model, the concurrent-gap condition is expected to break down. 
Under such circumstances, while the phase diagram in Fig.~\ref{fig:layeredRTP_phasediag}(\textbf{a}) would look qualitatively similar~\cite{Khalaf}, one should distinguish lines of surface gap closings vs.~lines of closing the Wannier gap at $\nu=\tfrac{1}{2}$. 
The presence of the metallic hinge mode (with a nontrivial hinge Berry phase) is governed by the closing of the surface energy gaps rather than by the closing of the Wannier gaps~\cite{Yang:2020}. 
This same phenomenology generalizes to the other models of DWIs presented in the next sections.

Finally, let us clarify the nature of the layered RTP model when $\alpha < 2$ and $\gamma_\textrm{-tra}/\lambda_\textrm{-ter}>1$, corresponding to the bottom-right region in Fig.~\ref{fig:layeredRTP_phasediag}(\textbf{a}) and labeled as DWI$'$.
In this region, the model in the wire geometry as specified earlier exhibits no hinge modes; nevertheless, the Wannier bands are still characterized by the non-trivial RTP displayed in Fig.~\ref{fig:layeredRTP_bands}(\textbf{c}).
Owing to the delicate topological invariant of the Wannier bands, we still regard the model as a delicate Wannier insulator. 
Its phenomenology is, in fact, analogous to the DWI at $\gamma_\textrm{-tra}/\lambda_\textrm{-ter}<1$; however, the appearance of the helical hinge modes requires the change of the $z$-boundary termination. 
Specifically, one needs to redefine the unit cell in the $z$-direction so that it is centered at the boundary of the formerly used unit cell (while maintaining unbroken unit cells at both $z$-terminated boundaries). 
Equivalently, we should include (or remove) a \emph{single} RTP layer at both $z$-terminated boundaries.
This behavior is reminiscent of the boundary-termination-dependent appearance of topological corner modes in the Benalcazar-Bernevig-Hughes model~\cite{Benalcazar} and of other BOTIs obtained via the layering construction~\cite{Khalaf}. 
More broadly, such termination-dependent appearance of topological boundary states is the characteristic feature of extrinsic HOTIs~\cite{Geier:2018}.
This dependence on the choice of the unit cell is also shared by the layered DWIs discussed in the following sections of the present work.

\section{Layering of delicate topological 1D chain}
\label{sec:layering1D}

The layered RTP insulator, a prime example of a DWI, was constructed by alternately stacking copies of the delicate-topological RTP insulator with pairwise canceling values of the RTP invariant.
We anticipate that a broader class of DWIs can be constructed by stacking other instances of delicate topological insulators. 
In the remainder of the manuscript, we discuss two concrete examples of such a construction: the stacking of delicate-topological 1D chains in the present section, and the stacking of delicate-topological Chern dartboard insulators in the subsequent Sec.~\ref{sec:layeringCDI}.

To proceed, we introduce in Sec.~\ref{sec:1D-chain-single} a toy model of a 1D chain with two orbitals per site, which is symmetric under space-time inversion ($\mathcal{PT}$) and space inversion ($\mathcal{P}$) symmetry.
We also summarize here the complete topological classification of 1D models with such symmetries.
Since the delicate topology associated with few-band models in this symmetry class has not been previously analyzed in detail, we provide in Appendices~\ref{sec:1d-chain-two-bands} and~\ref{sec:1d-chain-three-bands} the derivation of stable and delicate topological invariants which can arise in such 1D chains in the presence of two or more energy bands.
The results of the classification are summarized in Table~\ref{1D-chain-PT-P-classification}.
Subsequently, in Sec.~\ref{sec:deli-chain-layered} we perform the layering construction, stacking alternating layers of the delicate topological chains with pairwise canceling topology, obtaining a DWI in two dimensions.

\begin{table*}[t]
\centering
\begin{tabular*}{0.8\textwidth}{@{\extracolsep{\fill}}lll}
    \hline\hline
    ${}$        &  $\mathcal{PT}$ only  & $\mathcal{PT}$ and $\mathcal{P}$ \vspace{0.15cm}  \\   
    $N=2$ bands & $\mathbb{Z}$ [Eq.~(\ref{eqn:winding-number-def})] & $\mathbb{Z}$ [Eq.~(\ref{eqn:winding-number-def}); parity indicated by symmetry indicator in Eq.~(\ref{eqn:sym-ind})] \vspace{0.15cm} \\   
    $N\,{>}\,2$ bands & $\mathbb{Z}_2$ [Eq.~(\ref{eqn:Berry-phase-full-BZ})] & delicate (if mutually disjoint $\mathcal{P}$ eigenvalues): $\mathbb{Z}_2$ [Eq.~(\ref{eqn:Berry-phase-half-BZ})]     \\
    ${}$ & ${}$  & stable: $\mathbb{Z}_2$ [Eq.~(\ref{eqn:Berry-phase-full-BZ})] \\
    \hline\hline
\end{tabular*}
\caption{
Stable and delicate topological invariants characterizing
spinless 1D chains with space-time-inversion ($\mathcal{PT}$) symmetry with or without inversion ($\mathcal{P}$) symmetry depending on the number of bands (first column). 
The classification is discussed in Sec.~\ref{sec:1D-chain-single} and derived in Appendices~\ref{sec:1d-chain-two-bands} and~\ref{sec:1d-chain-three-bands}. 
The chain obeys the mutually disjoint condition if the $\mathcal{P}$ eigenvalue of all occupied bands at momenta $k_x\in\{0,\pi\}$ are opposite of the $\mathcal{P}$ eigenvalue of all the unoccupied bands at these momenta. 
}
\label{1D-chain-PT-P-classification}
\end{table*}

\subsection{Properties of a single delicate topological 1D chain}\label{sec:1D-chain-single}

We consider a one-dimensional chain with two orbitals per unit cell whose Bloch Hamiltonian is given by
\begin{align}
\mathcal{H}_{\textrm{chain}}(k_x) & =  h_x \sigma_x + h_z \sigma_z \nonumber \\
\textrm{with} \quad h_x(k_x) & = \sin(k_x) + \alpha  \sin(2 k_x) 
\label{eqn:1D-chain-double-wind}  \\
\textrm{and} \quad h_z(k_x) & = \cos(k_x) + \beta \cos( 2 k_x) \nonumber 
\end{align}
where $\alpha,\beta\in\mathbb{R}$ are tuning parameters.
To inspect the symmetry of the model, first observe that the reality of the Bloch Hamiltonian (at each $k_x$) implies the presence of space-time-inversion symmetry, represented as $(\mathcal{PT})\mathcal{H}_{\textrm{chain}}(k_x) (\mathcal{PT})^{-1}=\mathcal{H}_{\textrm{chain}}(k_x)$ with $\mathcal{PT} = \mathcal{K}$. 
Next, inversion symmetry $\mathcal{P} = \sigma_z$ is also present in the model: $\mathcal{P}\mathcal{H}_{\textrm{chain}}(k_x) \mathcal{P}^{-1}=\mathcal{H}_{\textrm{chain}}(-k_x)$, allowing us to characterize the constituent orbitals as even ($s$) resp.~odd ($p$) and to introduce a symmetry indicator.
Since the representation of $\mathcal{P}$ does not depend on $k_x$, it follows that both orbitals are situated at the center of the unit cell~\cite{Alexandradinata:2015}.
By combining the above two symmetries, the model is also symmetric under time reversal, $\mathcal{T} = \sigma_z \mathcal{K}$.
Finally, due to the absence of a  term proportional to the identity matrix $\sigma_0$, the Bloch Hamiltonian anticommutes with the chiral operator $\mathcal{S}=\sigma_y$.
However, such chiral symmetry is accidentally present in any traceless two-band model with $\mathcal{PT}=\mathcal{K}$~\cite{Takahashi:2024}.
Since we shall soon focus on a delicate topological invariant that persists in models with multiple bands, in which case this accidental chiral symmetry is lost, our subsequent discussion revolves only around the role of the $\mathcal{PT}$ and $\mathcal{P}$ symmetry.

Before analyzing the model in Eq.~(\ref{eqn:1D-chain-double-wind}), we summarize the stable and delicate topological invariants characterizing 1D chains with $\mathcal{P}$ and $\mathcal{PT}$ symmetry.
The problem can be approached either via homotopic characterization of the spectrally flattened Hamiltonian (which uses information from the full Brillouin zone $k_x\in[0,2\pi]$) or via symmetry indicators (which characterizes the Hamiltonian only at the high-symmetry momenta $k_x\in\{0,\pi\}$).
For two-band models, examined in Appendix~\ref{sec:1d-chain-two-bands}, the flattened $\mathcal{PT}$-symmetric Hamiltonians are expanded into the available Pauli matrices~as
\begin{equation}
\label{eqn:flattened-Hamiltonian}
\mathcal{H} = \left(h_x \sigma_x + h_z \sigma_z\right)/\sqrt{h_x^2 +h_z^2}.
\end{equation}
The classifying space of such Hamiltonians is a circle ($S^1$), meaning every Hamiltonian is a continuous map from the Brillouin zone circle into the circular classifying space. 
Such maps are distinguished by elements of the homotopy group 
\begin{equation}
\label{eqn:unstable-homotopy}
\omega \in \pi_1(S^1) = \mathbb{Z},
\end{equation}
where $\omega$ quantifies the winding number of the vector $\boldsymbol{h}=(h_x,h_z)$ around the origin $\boldsymbol{0}=(0,0)$ as the momentum $k_x$ sweeps along the 1D BZ.

The winding number can be computed from the original (i.e., non-flattened) Hamiltonian as 
\begin{equation}
\label{eqn:winding-number-def}
\omega = - \!\! \int_{-\pi}^{+\pi}  \!\!\! d k_x \frac{h_x \,\partial_{k_x} h_z - h_z \, \partial_{k_x} h_x}{h_x^2 + h_z^2}.
\end{equation}
The even (odd) values of $\omega$ correspond to a quantized Berry phase $\phi_\textrm{B}=0$ ($\phi_\textrm{B}=\pi$) in the 1D BZ, computed as
\begin{equation}
\label{eqn:Berry-phase-full-BZ}
\phi_\textrm{B}=\arg \left[\bra{u_\textrm{val}(2\pi)} \left( \ordprod_{k_x\in[0,2\pi]}\!\!\! \mathbb{P}_\textrm{val}(k_x)\right)\ket{u_\textrm{val}(0)}\right] 
\end{equation}
where $\mathbb{P}_\textrm{val}(k_x)=\ket{u_\textrm{val}(k_x)}\bra{u_\textrm{val}(k_x)}$ is the projector onto the valence band, $\ket{u_\textrm{val}(2\pi)}=\ket{u_\textrm{val}(0)}$ (due to the periodicity of the momentum space and the adoption of the periodic Bloch convention), and the symbol $\ordprod$ indicates a path-ordered product (with lower values of the the argument $k_x$ placed to the right).
The reality of $e^{i\phi_\textrm{B}}$ is guaranteed by $\mathcal{PT}$ symmetry.\footnote{Because the Hamiltonian is real at each $k_x$, a real gauge choice exists for which  $\ket{u_\textrm{val}(k_x)}$ is real-valued at each $k_x$. In such a real gauge, $\ket{u_\textrm{val}(2\pi)}=\pm\ket{u_\textrm{val}(0)}$, with $\pm 1$ being the phase holonomy, i.e., the Berry phase.}

It is further shown in the Appendix that the $\mathbb{Z}$-valued homotopic classification of two-band models is compatible with inversion symmetry (though the inversion implies that the full information about the Hamiltonian is contained in the half-BZ $k_x \in [0,\pi]$). 
In contrast, the only symmetry indicator is 
\begin{equation}
\label{eqn:sym-ind}
\xi_\textrm{val}(0)\xi_\textrm{val}(\pi)\in\{+1,-1\},    
\end{equation}
with $\xi_\textrm{val}(k)$ the inversion eigenvalue of the occupied (i.e., valence) band at the high-symmtry momentum $k_x$. 
The quantity~(\ref{eqn:sym-ind}) corresponds to a $\mathbb{Z}_2$ reduction of the $\mathbb{Z}$-valued winding number $\omega$.

Let us next enrich the model with a third band, which for concreteness we include in the conduction sector.
The classifying space is the real projective plane, $M=\mathbb{R} P^2$. 
First, in the absence of inversion symmetry, the homotopy group $\pi_1(\mathbb{R} P^2) = \mathbb{Z}_2$ exactly reproduces the $\mathbb{Z}_2$ characterization via the symmetry indicator in Eq.~(\ref{eqn:sym-ind}), or equivalently, by the quantized Berry phase $\phi_\textrm{B}$.
Therefore, one anticipates that the topological obstruction of two-band models with even $\omega$ (equivalently, with $\phi_\textrm{B}=0$) is trivialized by expanding the conduction subspace, i.e., that these 1D chains are DIs.

We further consider the inclusion of the inversion symmetry to the characterization of three-band models.
We show in Appendix~\ref{sec:1d-chain-three-bands} that such an inversion-symmetry-adapted description, carried in terms of \emph{relative} homotopy groups~\cite{Hatcher:2002,Sun:2018}, enhances the classification by one further topological class.
Namely, all three-band inversion-symmetric Hamiltonians with the nontrivial symmetry indicator ($\xi_\textrm{val}(0)\xi_\textrm{val}(\pi)=-1$) and with specified inversion eigenvalues of the occupied and unoccupied bands are homotopically equivalent; in contrast, Hamiltonians with \emph{trivial} symmetry indicator ($\xi_\textrm{val}(0)\xi_\textrm{val}(\pi)=1$) can be homotopically partitioned into two distinct classes if the three bands obey the mutually disjoint condition of inversion eigenvalues. 
The disjointness condition is fulfilled, for instance, if the occupied band is $s$-like at both $k_x \in\{0,\pi\}$ while the two unoccupied bands are both $p$-like at both momenta.
Of these two gapped Hamiltonians, one is not deformable to a unicellular atomic limit, i.e., it~is~a~DI.

This DI is distinguished from the trivial 1D chain by a $\mathbb{Z}_2$-quantized Berry phase $\phi_\textrm{half-B}$ over the half-BZ $k_x\in[0,\pi]$, defined as
\begin{equation}
\label{eqn:Berry-phase-half-BZ}
\phi_\textrm{half-B}=\arg \left[\bra{u_\textrm{val}(\pi)} \left( \ordprod_{k_x\in[0,\pi]}\!\!\! \mathbb{P}_\textrm{val}(k_x)\right)\ket{u_\textrm{val}(0)}\right]  
\end{equation}
where we have assumed a gauge that obeys $\ket{u_\textrm{val}(0)}=\ket{u_\textrm{val}(\pi)}$. 
Such gauge exists because the mutually disjoint arrangement of the inversion eigenvalues ensures $\mathbb{P}_\textrm{val}(0)=\mathbb{P}_\textrm{val}(\pi)$. 
Two-band models with winding number $\omega=4n+2$ are mapped to the nontrivial class with $\phi_\textrm{half-B}=\pi$, whereas two-band models with $\omega = 4n$ are mapped to the trivial class with $\phi_\textrm{half-B}=0$. 
This symmetry-protected $\mathbb{Z}_2$-valued delicate topology of 1D chains with trivial symmetry indicator persists in models with arbitrarily many bands (both in the conduction and in the valence band subspace) as long as the mutually disjoint condition of inversion eigenvalues is retained for all bands. 
In such many-band scenario, the Berry phase can be conveniently extracted from the path-ordered product 
\begin{equation}
\mathcal{W}_\textrm{val}^{\pi \leftarrow 0} = \ordprod_{k_x\in[0,\pi]}\mathbb{P}_\textrm{val}(k_x)
\end{equation}
as the complex argument of the product of unimodular (i.e., non-zero) eigenvalues of $\mathcal{W}_\textrm{val}^{\pi \leftarrow 0}$. 
The derived classification of 1D chains with $\mathcal{PT}$ and with or without $\mathcal{P}$ symmetry is summarized in Table~\ref{1D-chain-PT-P-classification}.

\begin{figure}
    \centering
    \includegraphics[width=0.99\linewidth]{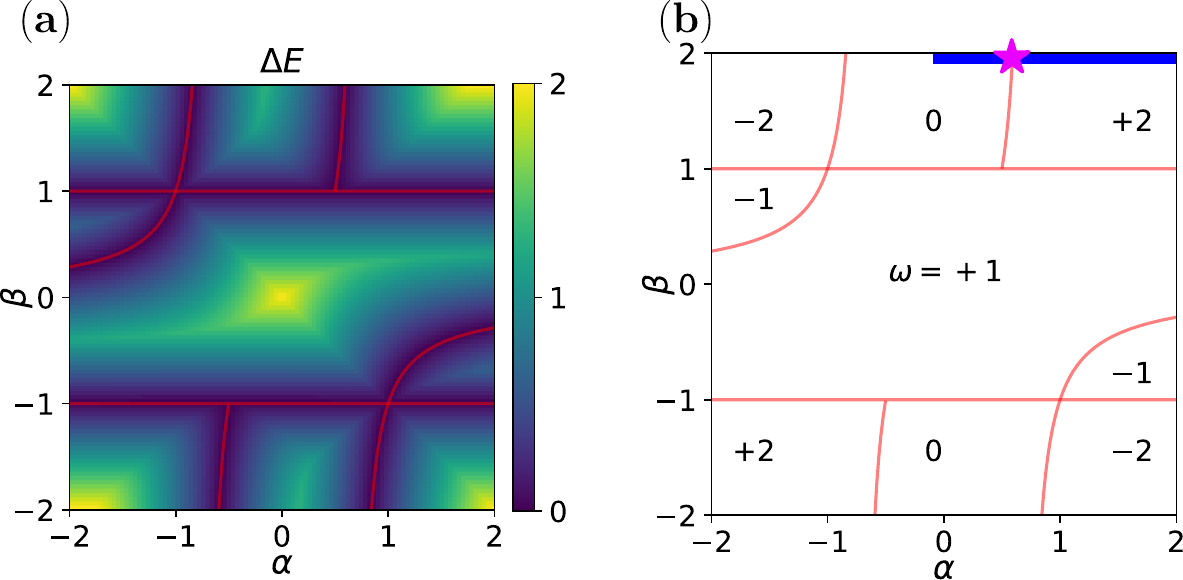}
    \caption{
    Phase diagram of the Hamiltonian $\mathcal{H}_{1\textrm{D}}$ in Eq.~(\ref{eqn:1D-chain-double-wind}) as a function of parameters $(\alpha,\beta)$. 
    (\textbf{a})~Size of the direct single-particle gap.
    The red lines indicate the values of ($\alpha$,$\beta$)
    where the bulk energy bands touch for some momentum inside the 1D BZ. 
    These lines can be obtained analytically [see Eq.~(\ref{eqn:1D-chain-phasediagram})], and they separate regions with different values of the winding number $\omega$. 
    (\textbf{b})~Winding number $\omega$ [Eq.~(\ref{eqn:winding-number-def})] of $\mathcal{H}_{1\textrm{D}}(k_x)$ inside the individual gapped regions of the phase diagram.
    For phases with $\omega$ an even number, Berry phase over half Brillouin zone can be computed [Eq.~(\ref{eqn:Berry-phase-half-BZ})]; namely $\phi_\textrm{half-B}=0$ for $\omega = 0$ and $\phi_\textrm{half-B}=\pi$ for $\omega=\pm 2$.
    The highlighted blue segment in the top-right corner of the phase diagram indicates the range $\alpha\in[0,2]$ and the value $\beta=2$, which are further considered in Figs.~\ref{fig:stacking_1d} and~\ref{fig:layered-chain-gaps}. 
    The purple star indicates the critical value $\alpha = \alpha_\textrm{c}$.
    }    
\label{fig:stacking_1d_phasediag}
\end{figure}

Having established the topological invariants available in one-dimensional two-band and multi-band models with $\mathcal{P}$ and $\mathcal{PT}$ symmetry,
we compute for the two-band model in Eq.~(\ref{eqn:1D-chain-double-wind}) the direct single-particle energy gap [Fig.~\ref{fig:stacking_1d_phasediag}(\textbf{a})] and the winding number $\omega$ [Fig.~\ref{fig:stacking_1d_phasediag}(\textbf{b})] as a function of parameters $\alpha, \beta \in[-2,2]$. 
The red lines in both panels of Fig.~\ref{fig:stacking_1d_phasediag} indicate parameters for which the bulk energy bands touch at some momentum inside the $1\textrm{D}$ BZ, indicating topological phase transitions.
Besides the horizontal phase-transition lines at $\beta = -1$ and $\beta = +1$, which correspond to closing of the energy gap at $k_x = 0$ resp.~at $k_x = \pi$, there are additional curved transition lines at
\begin{equation}
\label{eqn:1D-chain-phasediagram}
\beta = \alpha/(1-2\alpha^2)\quad\textrm{for}\quad\abs{\alpha}>1/2.
\end{equation}
The transition lines separate regions with winding numbers in the range $\omega\in\{-2,\ldots,+2\}$.

In the following discussion we fix $\beta = 2$ in the Hamiltonian in Eq.~(\ref{eqn:1D-chain-double-wind}), such that the occupied (unoccupied) band is $p$-like ($s$-like) at both $k_x\in \{0,\pi\}$.
We further restrict our attention to $\alpha>0$. According to the phase diagram in Fig.~\ref{fig:stacking_1d_phasediag}(\textbf{b}), we obtain a model with a single positive real parameter $\alpha$ that tunes between phases characterized by different winding numbers.
The winding number is $\omega = 0$ for $\alpha < \alpha_\textrm{c}$, and $\omega = 2$ for $\alpha > \alpha_\textrm{c}$, with the critical point [purple star in Fig.~\ref{fig:stacking_1d_phasediag}(\textbf{b})] determined from Eq.~(\ref{eqn:1D-chain-phasediagram}) as
\begin{equation}
\label{eqn:1D-alpha-crit}
    \alpha_\textrm{c} = (\sqrt{33}-1)/8 \approx 0.593.
\end{equation}
The winding number changes by $2$ at the critical point due to a pair of simultaneous closings of the energy gap at low-symmetry momenta $k_x$ and $-k_x$ related by the inversion symmetry.
This implies that the $\mathbb{Z}_2$-valued symmetry indicator in Eq.~(\ref{eqn:sym-ind}) (and correspondingly the $\mathbb{Z}_2$-quantized Berry phase over the 1D BZ) remains unchanged (namely trivial) during this topological band transition. 
In accordance with our earlier discussion, the phase with $\omega = 2$ exhibits a symmetry-protected delicate topology with $\phi_\textrm{half-B}=\pi$.

With sharp open boundaries, the topologically nontrivial nature of the 1D chain (with $2=\alpha>\alpha_\textrm{c}$) manifests as two boundary-localized zero-energy eigenstates shown in Fig.~\ref{fig:stacking_1d}(\textbf{a}). 
More generally, for a winding number $\omega$, there are $\abs{\omega}$ distinct eigenstates at each end of the system. 
These zero modes are pinned to $E = 0$ as long as the accidental chiral symmetry is present in the bulk and the edges are sharply terminated. 
If the bulk chiral symmetry is broken (e.g., through the addition of terms proportional to the identity $\sigma_0$ in the bulk Hamiltonian or by coupling to additional bands) or if the boundary termination is not sharp, the edge states generically move away from zero energy and are no longer energy-degenerate.

\subsection{Properties of the layered delicate topological 1D chain}\label{sec:deli-chain-layered}

We next consider stacking copies of the discussed $1$D model (with $\beta = 2$, and with tunable parameter $\alpha$) in the $y$-direction with both intralayer ($\gamma_\textrm{-tra}$) vs.~interlayer ($\lambda_\textrm{-ter}$) couplings, in the spirit of the layered RTP model illustrated in Fig.~\ref{fig:layeredRTP_construction}. 
The resulting Hamiltonian is expressed as 
\begin{eqnarray}
\label{eqn:layered-chain-model}
\mathcal{H}_{\,\textrm{chain}}^{\mathrm{layered}}(k_x,k_y)  &=&  \tau_z \otimes \mathcal{H}_{\textrm{chain}}(k_x)  + \lambda_\textrm{-ter} \sin k_y \, \tau_y \otimes \sigma_0  + \nonumber \\ 
&\phantom{=}& \quad
+ (\gamma_\textrm{-tra} + \lambda_\textrm{-ter} \cos k_y) \, \tau_x \otimes \sigma_0 .
\end{eqnarray}
The constructed Hamiltonian exhibits a mirror symmetry $M_x$, inherited from the inversion symmetry ($\mathcal{P}$) of the component 1D chains, which acts as 
\begin{equation}
\label{eq:layered-chain_mirror-symmetry}
    M^{-1}_x \mathcal{H}_{\,\textrm{chain}}^{\mathrm{layered}}(\boldsymbol{k}) M_x = \mathcal{H}_{\,\textrm{chain}}^{\mathrm{layered}}(M_x\boldsymbol{k}),
\end{equation}
with mirror-symmetry operator $M_x = \tau_0\otimes \sigma_z$ and with mirror-related momentum $M_x \boldsymbol{k} = (-k_x,k_y)$.

\begin{figure}
    \centering
    \includegraphics[width=0.99\linewidth]{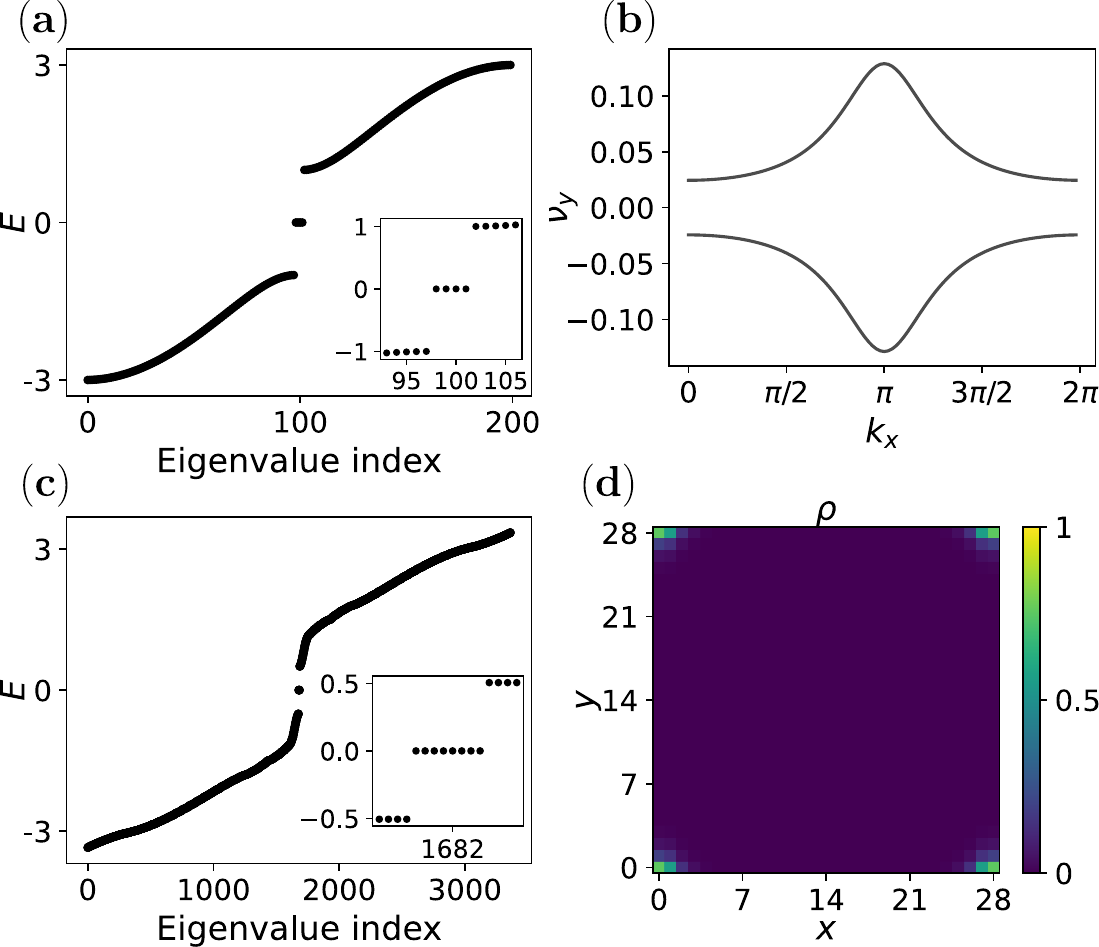}
    \caption{
    Layering of the delicate topological 1D chains.
    (\textbf{a})~The spectrum of the one-dimensional Hamiltonian $\mathcal{H}_{\textrm{chain}}$ in Eq.~(\ref{eqn:1D-chain-double-wind}) for $\alpha=2=\beta$ and  $N_x =100$ unit cells in the presence of sharp open boundaries.
    The inset magnifies the spectrum near $E=0$, revealing four boundary-localized eigenstates (two on each end). 
    (\textbf{b})~The Wannier bands in the $y$-direction computed for the stacked two-dimensional model in Eq.~(\ref{eqn:layered-chain-model}) for $\alpha=\beta=2$, $\gamma_\textrm{-tra} = 0.5$, and $\lambda_\textrm{-ter} =1$.
    (\textbf{c})~For the same parameters, and adopting the linear system size $N_x = N_y = 29$, we compute the spectrum of the layered model in flake geometry (sharp OBCs in both $x$ and $y$-directions). 
    The inset magnifies the spectrum near $E=0$, highlighting eight zero-energy eigenstates.
    (\textbf{d})~The cumulative probability density of the eight zero-energy eigenstates. 
    The symmetric distribution of the probability density implies two exponentially-localized zero-energy eigenstates at each corner of the flake.
    }
    \label{fig:stacking_1d}
\end{figure}

As a result of the layering construction we obtain two hybrid Wannier bands per unit cell with spectral eigenvalues $\nu_y^{\pm}(k_x)$ that are gapped for all $k_x$, as shown in Fig.~\ref{fig:stacking_1d}(\textbf{b}). 
By design, the wave functions of these Wannier bands have the same symmetry representations at high-symmetry momenta $k_x\in\{0,\pi\}$ as the energy bands of $\mathcal{H}_{\textrm{chain}}(k_x)$, implying that they carry trivial Berry phase, labeled $\phi_\textrm{B}^{\nu^y_\pm}=0$. 
Furthermore, since the Wannier bands also inherit the mutually disjoint condition with respect to the eigenvalues of $M_x$, they can be topologically characterized in terms of the same delicate invariants as the energy bands of $\mathcal{H}_{\textrm{chain}}(k_x)$.
In the following, we choose the description in terms of the Berry phase of the lower Wannier band over the half-BZ $k_x \in [0,\pi]$, denoted $\phi_\textrm{half-B}^{\nu^y_-}$. 
Due to the layering construction, the complementary (i.e., upper) Wannier band carries Berry phase $\phi_\textrm{half-B}^{\nu^y_+} = - \phi_\textrm{half-B}^{\nu^y_-}$;\footnote{This relation follows from the reflection symmetry 
\begin{equation}
M_y^{-1} \mathcal{H}^\textrm{layered}_\textrm{chain}(k_x,k_y) M_y = \mathcal{H}(k_x,-k_y),    
\end{equation} 
represented by $M_y=\tau_x\otimes \sigma_y$, which arises when the 1D chain Hamiltonian in Eq.~(\ref{eqn:1D-chain-double-wind}) does not contain a term proportional to the identity matrix $\sigma_0$.} however, the sign reversal is not important, due to the quantization of the Berry phases to $0$ vs.~$\pi$ (defined modulo $2\pi$) imposed by $M_x$.
This phase is computed using the formula in Eq.~(\ref{eqn:Berry-phase-half-BZ}) with the replacement [including inside the projectors $\mathbb{P}(k_x)$] of the valence energy state $\ket{u_\textrm{val}(k_x)}$ by the Wannier state $\ket{\nu^y_-(k_x)}$, and it is quantized by the mirror symmetry $M_x$ to two possible values: $0$ and $\pi$.

In analogy with the discussion of the layered RTP model, we find that the Berry phase $\phi_\textrm{half-B}^{\nu^y_-}$ of the lower Wannier band in $\mathcal{H}_{\,\textrm{chain}}^{\mathrm{layered}}$ matches exactly the Berry phase $\phi_\textrm{half-B}$ of the valence band in $\mathcal{H}_{\,\textrm{chain}}$. 
Specifically, tuning the parameter $\alpha/\alpha_\textrm{c}$ [with $\alpha_\textrm{c}$ specified in Eq.~(\ref{eqn:1D-alpha-crit})] allows us to control the delicate Wannier invariant, while the ratio of $\gamma_\textrm{-tra}/\lambda_\textrm{-ter}$ controls the dimerization pattern of the alternating layers.
It follows that the phase diagram of $\mathcal{H}_{\,\textrm{chain}}^{\mathrm{layered}}$, shown in Fig.~\ref{fig:layeredRTP_phasediag}(\textbf{b}), is similar to that of the layered RTP insulator [Fig.~\ref{fig:layeredRTP_phasediag}(\textbf{a})], in the sense that tuning either $\alpha/\alpha_\textrm{c}$ or the ratio $\gamma_\textrm{-tra} / \lambda_\textrm{-ter}$ can induce a transition through an energy gap closing at the one-dimensional edges.
The phases at $\alpha/\alpha_\textrm{c}>1$ are associated with a delicate invariant of the Wannier bands, i.e., they are DWIs.

\begin{figure}
    \centering
    \includegraphics[width=0.99\linewidth]{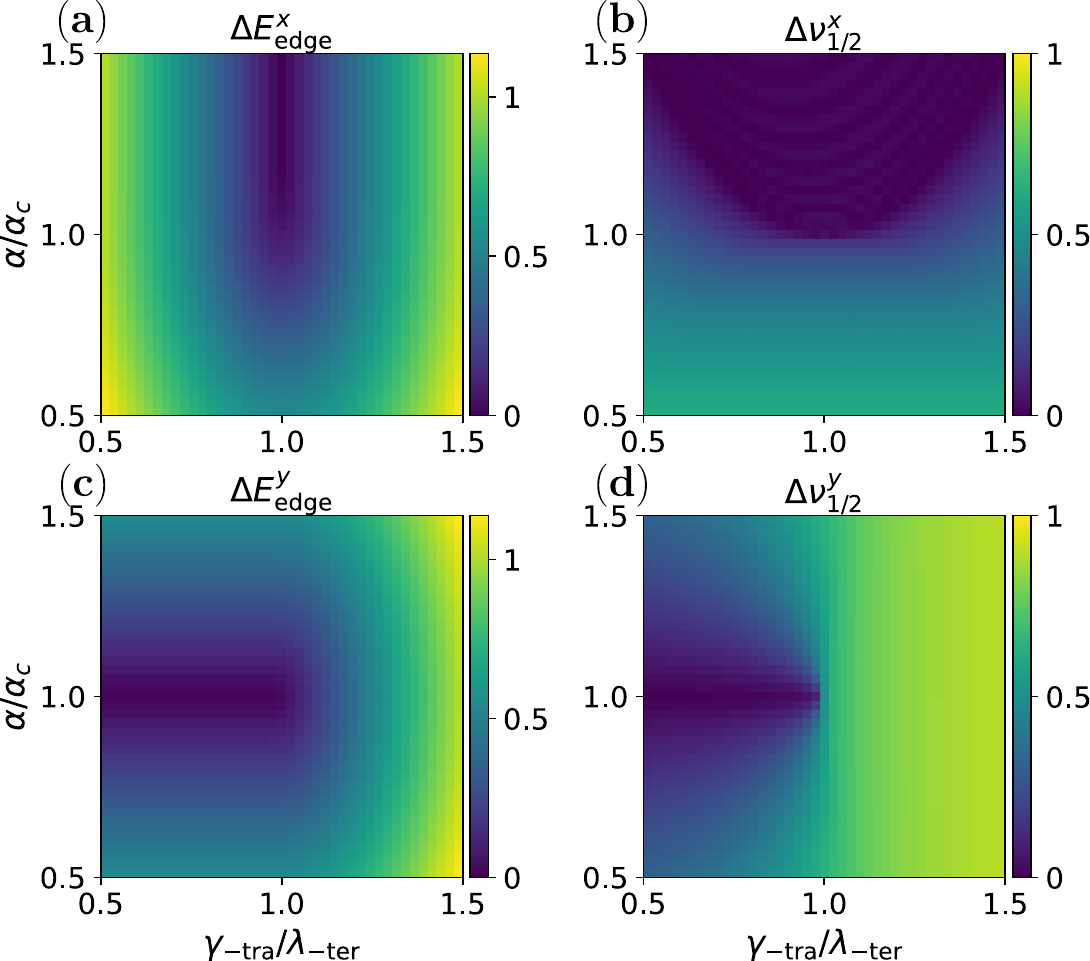}
    \caption{
    Spectral gaps of the layered chain model [Eq.~(\ref{eqn:layered-chain-model})] in ribbon geometry. 
    (\textbf{a})~The edge energy gap $\Delta E^x_\textrm{edge}$ of a ribbon with sharp open boundaries in the $x$-direction. 
    The gap closes at $\gamma_\textrm{-tra}/\lambda_\textrm{-ter}=1$ for $\alpha/\alpha_\textrm{c}>1$.
    (\textbf{b}) The closing of the energy gap is mimicked by the closing of the gap $\Delta \nu_{1/2}^x$ in the $x$-direction, although additional crossings of the Wannier bands at $\nu=1/2$ result in broadening of the gapless region.
    Panels (\textbf{c},\textbf{d}) display the analogous data for the $y$-direction. 
    Both gaps $\Delta E^y_\textrm{edge}$ and $\Delta \nu^y_{1/2}$ close at $\alpha/\alpha_\textrm{c} = 1$ for $\gamma_{\textrm{-tra}}/\lambda_{\textrm{-ter}}<1$.
    The data in (\textbf{a},\textbf{c}) is obtained for a system with $N_1 = 20$ layers and with $N_2 = 100$ momenta inside the ribbon Brillouin zone (BZ). 
    The data in (\textbf{b},\textbf{d}) is obtained from a grid of $N_x = 100 = N_y$ momenta inside the bulk BZ.
    We fixed $\beta=2$ and $\lambda_\textrm{-tra}=1$ throughout.
    }
    \label{fig:layered-chain-gaps}
\end{figure}

The closing of the edge energy gaps in the layered model~(\ref{eqn:layered-chain-model}) is correlated with the closing of the Wannier gap at $\nu = \tfrac{1}{2}$ in the corresponding direction.
To illustrate the correlation, we plot in Fig.~\ref{fig:layered-chain-gaps} the edge energy gaps [panels (\textbf{a},\textbf{c})] and the Wannier gaps at $\nu = \tfrac{1}{2}$ [panels (\textbf{c},\textbf{d})] in the $x$- and in the $y$-direction for $\beta = 2$ and $\lambda = 1$.
Within human-eye precision, the gap closings of $E_\textrm{edge}^y$ [Fig.~\ref{fig:layered-chain-gaps}(\textbf{b})] and $\nu^y_{1/2}$ [Fig.~\ref{fig:layered-chain-gaps}(\textbf{d})] are concurrent.
On the other hand, the gap closings of $E_\textrm{edge}^x$ [Fig.~\ref{fig:layered-chain-gaps}(\textbf{a})] along a line segment translates at the level of $\nu^x_{1/2}$ [Fig.~\ref{fig:layered-chain-gaps}(\textbf{c})] into a somewhat broadened gapless region due to the appearance of additional Dirac-type touchings of the Wannier bands at $\nu = \tfrac{1}{2}$. 
In other words, the concurrent-gap condition applies in the $y$-direction but fails in the $x$-direction (although the implication continues to apply in one way; namely, vanishing of the gap $\Delta E^x_\textrm{edge}$ implies vanishing of the gap $\Delta \nu^x_{1/2}$). 
Where the two critical lines meet (i.e., at $\alpha/\alpha_\textrm{c} = 1 = \gamma_\textrm{-tra}/\lambda_\textrm{-ter} = 1$), the model exhibits closing of the bulk energy gap.

To reveal the bulk-boundary correspondence of the layered Hamiltonian in Eq.~(\ref{eqn:layered-chain-model}) in the topological phase ($\alpha/ \alpha_\textrm{c} >1$, $\gamma_\textrm{-tra}/\lambda_\textrm{-ter} < 1$), we consider the flake geometry with sharply terminated edges in the $x$ and $y$-directions and with linear dimensions $N_x = N_y = 29$.
We diagonalize the corresponding Hamiltonian for parameter values $\alpha = \beta = 2$, $\gamma_\textrm{-tra} = 0.5$, and $\lambda_{\textrm{-ter}}=1$. 
The obtained spectrum, shown in Fig.~\ref{fig:stacking_1d}(\textbf{c}), exhibits eight near-zero-energy eigenstates.
To confirm that the eigenstates are indeed localized to the corners of the system, we plot in Fig.~\ref{fig:stacking_1d}(\textbf{d}) the cumulative probability density of the eight zero-energy eigenstates. 
As anticipated, the obtained distribution is concentrated at the corners of the flake, confirming the higher-order localization of these zero-energy states.
Switching to $\gamma_\textrm{-tra}/\lambda_\textrm{-ter}>1$ (while keeping $\alpha/\alpha_\textrm{c}>1$; not plotted) preserves the delicate topology of the Wannier bands (i.e., the system remains a DWI) while removing the zero-energy corner modes through an energy gap closing at the $x$-terminated edges.
In this regime, the corner modes associated with the delicate Wannier topology appear for an alternative choice of the $y$-termination, in a spirit analogous to the discussion of the layered RTP model in Sec.~\ref{sec:layered-RTP-PD&obst}.

The two-dimensional model defined in Eq.~(\ref{eqn:layered-chain-model}) has a boundary-obstructed character~\cite{Khalaf}, which can be understood in analogy with the discussion of the layered RTP insulator as follows.
In the absence of edges, one can move across the critical lines $\alpha/\alpha_\textrm{c}=1$ and $\gamma_\textrm{-tra}/\lambda_\textrm{-ter} = 1$ without closing the energy gap; these critical lines are only indicative of closing the \emph{edge} energy gaps. Therefore, the system can be trivialized (i.e., continuously deformed to a unicellular atomic limit) without closing the energy gap. 
The same conclusions remain true in the presence of OBC in one direction and with PBC in the other direction, i.e., in ribbon geometry. 
This is because in such geometry only one of the two energy-gap-closing lines [cf.~data in Fig.~\ref{fig:layered-chain-gaps}(\textbf{a}) resp.~(\textbf{c})] arises in the system, meaning that one can circumvent the energy gap closing and connect any choice of parameters inside the parameter plane.
In contrast, for a flake geometry with OBC in \emph{both} directions, the phase at $\alpha/\alpha_\textrm{c}>1$ and $\gamma_\textrm{-tra}/\lambda_\textrm{-ter} < 1$ is energetically separated from the phase at other values of these parameters. 
For the specified choice of sharp open boundary termination, this phase is characterized by the presence of two zero-energy modes at each corner,
which cannot be removed from the spectrum without closing the energy gap in the system (at least if assuming the approximate chiral symmetry of the spectrum).

\section{Layering of Chern dartboard insulator}
\label{sec:layeringCDI}

As our final example, we briefly discuss another case of a DWI in three spatial dimensions: the layered Chern dartboard insulator, which has a bulk invariant equal to a quantized Berry flux of the hybrid Wannier band in one quarter of the Brillouin zone. 
In the simplest model realization of this quantized flux, the insulator exhibits a pair of helical modes on hinges for two distinct orientations of a wire.
However, in contrast to the discussion of the layered RTP model, the bulk-hinge correspondence of this DWI is not robust even if narrowing our attention to sharp boundary terminations: the hinge modes can be gapped by perturbing the \emph{bulk} Hamiltonian and the hinges are not associated with an anomalous hinge Berry phase.
In other words, while the discussed model is a DWI, its topological invariant does not appear to be associated with robust higher-order or boundary-obstructed signatures.
This observation explicitly demonstrates that the concept of DWIs is distinct from both HOTIs and BOTIs.

Our discussion is structured as follows.
First, we present in Sec.~\ref{sec:single-CDI-layer} a concrete model of a Chern dartboard insulator (CDI). 
This is a two-dimensional insulator whose delicate topological invariant is a sub-Brillouin zone Chern number protected by a collection of mirror planes.
We discuss the stability of this topological invariant and consider its bulk-edge correspondence.
In Sec.~\ref{sec:CDI-layered}, we apply the layering construction, based on stacking alternating layers of CDIs with pairwise canceling topology, to obtain the layered CDI in three dimensions.
We find that the layering construction elevates the delicate topology of energy bands of a single CDI layer to delicate topology of Wannier bands of the layered model, and translates the edge spectrum of the single CDI layer to the hinge spectrum of the layered model.

\subsection{Properties of a single Chern-dartboard layer}\label{sec:single-CDI-layer}

We adapt from Ref.~\onlinecite{Chen_2024} the model of an $n=2$ \emph{Chern dartboard Insulator} ($\mathrm{CDI}_2$). Its Bloch Hamiltonian takes the form
\begin{align}
\label{eqn:CDI-Ham}
\mathcal{H}_{\mathrm{CDI}_2\!}(k_x,k_y) \!&=\! -\! \sin(k_x) \sin(2 k_y) \sigma_x +\! \sin(2 k_x) \sin(k_y) \sigma_y + \nonumber \\
&\phantom{=} + \!\left( \alpha + \cos(2 k_x) + \cos(2 k_y) \right) \sigma_z.
\end{align}
The model has two orthogonal mirror symmetries, $M_x$ and $M_y$, both represented as $\sigma_z$.
Such symmetry is consistent with placing one $s$ and one $d_{xy}$ orbital in the middle of each unit cell.
Focusing our attention on positive values of the tuning parameter $\alpha$, the model exhibits two insulating phases: delicate topological ($0 < \alpha < 2$) and trivial ($\alpha > 2$).

\begin{figure}
    \centering
    \includegraphics[width=0.99\linewidth]{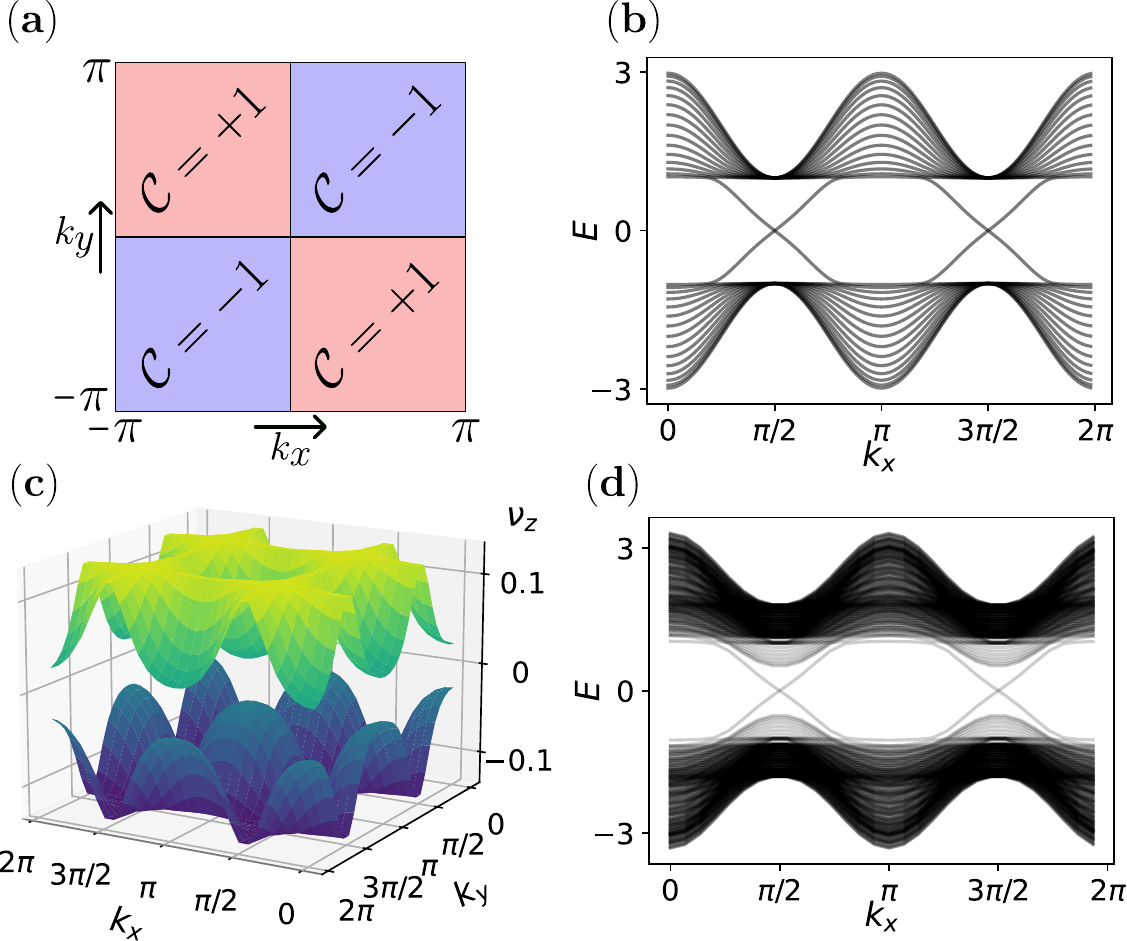}
    \caption{
    Layering of the $n=2$ Chern dartboard insulator ($\mathrm{CDI}_2$).  
    (\textbf{a})~Arrangement of sub-BZ Chern numbers across the Brillouin zone quadrants for the occupied band of the $\textrm{CDI}_2$ Hamiltonian in Eq.~(\ref{eqn:CDI-Ham}) in the topological phase with $\alpha = 1$.
    (\textbf{b})~Spectrum of $\textrm{CDI}_2$ in ribbon geometry with sharp OBC in the $y$-direction for $\alpha = 1$ and $N_y = 30$. 
    Each open edge exhibits a pair of counter-propagating metallic edge states connecting bulk conduction and valence bands.
    These metallic edge states can be gapped, e.g., by the perturbation in Eq.~(\ref{eqn:CDI2-gapped-edges}).
    (\textbf{c})~The Wannier bands of the layered $\textrm{CDI}_2$ in Eq.~(\ref{eqn:layered-CDI-Ham}) with  $\alpha = 1$, $\gamma_\textrm{-tra} = 0.5$, and $\lambda_\textrm{-ter} = 1$.
    We numerically verify that the Wannier bands mimic the arrangement of the sub-BZ Chern numbers of $\mathcal{H}_{\textrm{CDI}_2}$ in panel~(\textbf{a}).  
    (\textbf{d})~Energy spectrum of the layered $\textrm{CDI}_2$ insulator in the wire geometry with sharp OBC in the $y$ and $z$-direction and PBC in the $x$-direction.
    We used the same Hamiltonian parameters as in panel (\textbf{c}) and wire dimensions $N_y = N_z = 20$. 
    Each band connecting conduction to valence is fourfold degenerate, indicating a pair of helical metallic modes localized at every hinge of the wire.
    }
    \label{fig:stacking_2d}
\end{figure}

The topological phase of $\mathrm{CDI}_2$ exhibits a ``dartboard'' pattern of quantized Berry fluxes across quadrants of the BZ as shown in Fig.~\ref{fig:stacking_2d}(\textbf{a}). 
The topological robustness of the Berry curvature integrals over each BZ quadrant is protected by the mirror symmetries, which guarantee that the occupied Bloch wave function is constant (being either even or odd under the reflection symmetries) along all high-symmetry lines $k_{x,y}\in\{0,\pi\}$, which  form a rectangular grid over the 2D BZ.
For this reason, the boundary of each BZ quadrant can be identified as a single point~\cite{Sun:2018}, effectively turning the quadrant into $S^2$.
The $\mathbb{Z}$-valued Chern number on the BZ quadrant $[0,\pi]\times[0,\pi]$, which we further refer to as the \emph{sub-BZ Chern number}, is a delicate topological invariant, which obstructs the construction of unicellular symmetric Wannier orbitals~\cite{Nelson_prl}. 
The delicate aspect of the invariant is revealed as follows: inclusion of additional trivial bands that violate the mutually disjoint $M_x$ or $M_y$ mirror eigenvalues of the valence vs.~conduction sector allows the Bloch wave function to vary along the high-symmetry lines. 
This variation prevents us from compactifying the BZ quadrant into a closed surface, which results in the loss of the quantization of the Berry flux.

It was reported in Ref.~\onlinecite{Chen_2024}, based on numerical observations, that the nonzero sub-BZ Chern number of the phase at $\alpha = 1$ is manifested by the appearance of metallic edge states in ribbon geometry with sharp OBC.
We reproduce these edge states for a ribbon infinite in the $x$-direction and with $N_y = 30$ unit cells in the $y$-direction in Fig.~\ref{fig:stacking_2d}(\textbf{b}).
Owing to the symmetric arrangement of the quantized Berry fluxes, a ribbon infinite in the $y$-direction and with sharp open boundaries in the $x$-direction exhibits the same spectrum features.
In fact, for the model in Eq.~(\ref{eqn:CDI-Ham}), the two orientations of the ribbon are related by the magnetic symmetry 
\begin{equation}
\label{eqn:CDI2-magnetic-symmetry}
C_{4z}\mathcal{T}: (k_x,k_y) \mapsto (-k_y,k_x),
\end{equation}
represented by $\tfrac{1}{\sqrt{2}}(\sigma_0 - i \sigma_z)$, where $\mathcal{K}$ indicates complex conjugation.

The computed spectrum of the CDI$_2$ ribbon is reminiscent of the spectrum of the RTP ribbon, displayed in Fig.~\ref{fig:basicRTP}(\textbf{b}).
However, there is an important distinction; namely, the multiplicity of the helical edge states is \emph{doubled} when compared to the RTP insulator. 
The doubling implies that every spectral branch that appears in Fig.~\ref{fig:stacking_2d}(\textbf{b}) inside the bulk energy gap is twofold degenerate, with one copy of these in-gap states at both edges.
In particular, each edge exhibits a pair of metallic branches that cross at momenta $k_x = \pm \pi/2$ at energy $E=0$.\footnote{The two-fold energy-degenerate zero modes can be understood from $\mathcal{H}_{\mathrm{CDI}_2\!}(\pi/2,k_y)=-\sin(2k_y)\sigma_x+\cos(2k_y)\sigma_z$ being chiral symmetric [$\sigma_y\mathcal{H}_{\mathrm{CDI}_2\!}(\pi/2,k_y)\sigma_y=-\mathcal{H}_{\mathrm{CDI}_2\!}(\pi/2,k_y)$] with winding number two. The perturbation  in Eq.~(\ref{eqn:CDI2-gapped-edges}) breaks this chiral symmetry.}
However, these crossings are not protected by the mirror symmetries, even if restricting our attention to sharp boundary conditions.
Indeed, it is easily verified that the bulk perturbation 
\begin{equation}
\label{eqn:CDI2-gapped-edges}
\mathcal{H}_\textrm{pert}\propto (\cos k_x + \cos k_y)\sigma_0,  
\end{equation}
which preserves both mirror symmetries as well as the quantization of the sub-BZ Chern number, results in hybridization of the edge states, thus opening an energy gap at the edges.\footnote{
Whether CDI$_2$ exhibits an actually robust bulk-boundary correspondence remains at present an open question.
}
In the following discussion, we adopt the original CDI$_2$ model of Ref.~\onlinecite{Chen_2024} without perturbations.
While such a model exhibits edge states in ribbons with sharp open boundaries, one should bear in mind that these edge states appear only due to fine-tuning of the Hamiltonian parameters.

\subsection{Properties of the layered Chern dartboard insulator}\label{sec:CDI-layered}

We elevate the nontrivial texture of Berry curvature from energy bands to Wannier bands by stacking in the $z$-direction alternating layers of $\mathrm{CDI}_2$ with pairwise canceling value of the sub-BZ Chern number.
Following the schematics in Fig.~\ref{fig:layeredRTP_construction}, we use $\gamma_\textrm{-tra}$ for the intralayer coupling and $\lambda_\textrm{-ter}$ for the interlayer coupling.
The resulting layered $\textrm{CDI}_2$ Hamiltonian is
\begin{eqnarray}
\label{eqn:layered-CDI-Ham}
\mathcal{H}_{\,\mathrm{CDI}_2}^{\mathrm{layered}}(\boldsymbol{k}) &=&  \tau_z \otimes \mathcal{H}_{\mathrm{CDI}_2}(k_x,k_y)  + \lambda_\textrm{-ter} \sin k_z \, \tau_y \otimes \sigma_0 + \nonumber \\ 
&\phantom{=}& \quad + (\gamma_\textrm{-tra} + \lambda_\textrm{-ter} \cos k_z) \, \tau_x \otimes \sigma_0.
\end{eqnarray}
The Wannier bands in the $z$-direction computed for $\alpha = 1$, $\gamma_\textrm{-tra} = 0.5$, $\lambda_\textrm{-ter} = 1$ are shown in Fig.~\ref{fig:stacking_2d}(\textbf{c}).
To reveal the nontrivial profile of quantized Berry fluxes, we first find eigenstates $\ket{\nu_z^\pm(k_x,k_y)}$ of the projected position operator in the $z$-direction. 
Then, we compute the subBZ Chern number for the lower Wannier band $\ket{\nu_z^-}$ inside the region $[0,\pi] \times [0,\pi]$.
Our numerical computation~\cite{supp} finds $\mathcal{C}=-1$, which corresponds to the subBZ Chern number of $\mathcal{H}_{\mathrm{CDI}_2}$ by virtue of the layering construction.
Since the Wannier bands exhibit a delicate topological invariant, the layered CDI$_2$ model~is~a~DWI.

As the next step, we numerically investigate the hypothetical bulk-hinge correspondence of the layered $\textrm{CDI}_2$ model. 
We specifically consider a wire geometry with sharp OBC in the $y$ and $z$-directions, while $k_x$ remains a good quantum number.\footnote{Due to the magnetic $C_{4z}\mathcal{T}$ symmetry, given by Eq.~(\ref{eqn:CDI2-magnetic-symmetry}), an equivalent discussion applies to a wire infinite in the $y$-direction with sharp OBC in the $x$ and $z$-directions.}
The obtained spectrum, plotted as a function of $k_x$, is shown in Fig.~\ref{fig:stacking_2d}(\textbf{d}).
As anticipated, the layering construction elevates the metallic edge modes of the CDI$_2$ to metallic hinge modes of the layered CDI$_2$ model.
Specifically, each spectral branch located around $E=0$ in Fig.~\ref{fig:stacking_2d}(\textbf{d}) is found to be \emph{fourfold} degenerate.
The additional doubling (in addition to the doubling previously discussed for a single CDI$_2$ layer in Sec.~\ref{sec:single-CDI-layer}) appears as a fingerprint of the bottom and the top CDI$_2$ layer of the wire.
Each of the four hinges exhibits two counterpropagating metallic modes that intersect at $k_x = \pm \pi/2$ and~$E=0$.

In analogy with the discussion of the single CDI$_2$ layer, the zero-energy crossing of the hinge modes in Fig.~\ref{fig:stacking_2d}(\textbf{d}) is not topologically protected.
In particular, inclusion of the perturbation in Eq.~(\ref{eqn:CDI2-gapped-edges}) within the layered Hamiltonian results in hybridization of the metallic modes, thus opening a spectral gap at the hinges.
In addition, while the layered RTP insulator has a nontrivial hinge Berry phase, which remains well-defined even after removing the metallic hinge modes with a local perturbation, the layered dartboard insulator has a trivial hinge Berry phase.
The available evidence suggests that, in contrast with the DWIs discussed in Secs.~\ref{sec:layered_rtp} and~\ref{sec:deli-chain-layered}, the delicate Wannier topology of the layered CDI$_2$ model is not associated with a recognizable robust boundary signature.

\section{Conclusions and outlooks}
\label{sec:conclusion}

In this work, we introduced a refined class of topological insulators (TIs), dubbed \emph{delicate Wannier insulators} (DWIs). Their defining principle is the elevation of a delicate topological invariant from Bloch bands (energy eigenstates) to Wannier bands (eigenstates of a projected position operator).
The resulting Hamiltonians are gapped in the bulk and on the first-order boundaries (i.e., edges or surfaces) but they may exhibit topologically robust metallic states on sharply terminated higher-order boundaries (i.e., hinges or corners).
DWIs can be deformed without closing the energy gap to the unicellular atomic limit (i.e., with symmetric Wannier orbitals contained within a single unit cell) in the bulk as well as in the presence of a single open boundary.
In contrast, trivialization of the Wannier topology may be obstructed by energy gap closing in the presence of two sharply terminated open boundaries that meet at a hinge or a corner. 
Specifically, such obstruction applies in DWIs that adhere to the concurrent-gap condition, i.e., where closing of the energy gap at sharp open surfaces conincides with the closing of the Wannier gap at $\nu = 1/2$ in the corresponding direction.
For sufficiently generic models, the concurrent-gap condition is expected to break down, resulting in a more subtle structure of the phase diagrams.

To usher this refined class of TIs, we illuminated the key underlying concepts and properties with the paradigm example of a DWI in three spatial dimensions: the layered RTP insulator.
This model is obtained by coupling alternating layers of two-dimensional delicate-topological RTP insulators with pairwise canceling values of their topological invariant.
The layering construction~\cite{Khalaf} ensures that the layered RTP insulator in wire geometry exhibits metallic helical modes at the hinges, that are reminiscent of the helical edge modes of a single RTP-insulating layer.
Further examples of $(d+1)$-dimensional DWIs can be generated by applying the layering construction to delicate topological insulators (DIs) in $d$ dimensions, such as the ones mathematically characterized in Ref.~\onlinecite{Brouwer}.
We illustrated the general principle with two concrete examples: layering of a delicate-topological 1D chain (which results in a DWI in two dimensions) and layering of a two-dimensional Chern dartboard insulator from Ref.~\onlinecite{Chen_2024} (which results in a DWI in three dimensions)   .

The phenomenology of DWIs is distinct from the formerly introduced categories of TIs.
In particular, their topology is inivisible to symmetry indicators characterizing the bulk valence bands (in contrast to stable and fragile TIs~\cite{Po:2017}), to the homotopy characterization of the bulk Bloch Hamiltonian (in contrast to delicate TIs~\cite{Brouwer}), and even to symmetry eigenvalues of the Wannier bands (in contrast to BOTIs~\cite{Khalaf}).
The triviality of the symmetry indicators implies that DWIs are Wannierizable, and the triviality of the homotopy equivalence class means that the symmetric Wannier orbitals can be made unicellular upon a continuous gap- and symmetry-preserving deformation of the bulk Hamiltonian.
Notably, since the topological invariant is extracted from the Wannier spectrum of the valence bands, the topology is unaffected by inclusion of additional bands to the conduction sector.
From this perspective, the robustness of DWIs is enhanced compared to that of delicate topological insulators (DIs) and is more reminiscent of fragile topological insulators (FIs).
To the extent that concurrent-gap condition holds in certain elementary models, DWIs also have the property of boundary obstruction against unicellularity (similar in spirit to the obstruction in BOTIs). 
Finally, while all DWIs have well-defined bulk invariants protected by the Wannier gap, only some DWIs have robust higher-order modes (i.e., some though not all DWIs are HOTIs). 
For example, we found that the layered RTP insulator has a robust hinge Berry phase, whereas the layered CDI model appears to have no robust higher-order signature.

While our work presents the basic phenomenology of DWIs, we were unable to support certain observations with rigorous mathematical proofs and rely instead on numerical evidence. 
Therefore, further research into DWIs is necessary to bring some of the discussed properties on a solid footing.
Below, we comment on several concrete directions that would benefit from a dedicated mathematical analysis and follow-up studies.

First, consider again the higher-order bulk-boundary correspondence of DWIs. 
For concreteness, we focus here on the layered RTP insulator, for which we observe the appearance of metallic states at sharply terminated hinges.
While the appearance of metallic states for delicate-topological RTP insulator at first-order boundaries was proved using spectral properties of block Toeplitz matrices~\cite{Nelson_prb,Miranda:2000}, generalization of this approach to higher-order boundaries of DWIs would require a higher-dimensional generalization of block Toeplitz matrices.
As we are at present not familiar with spectral properties of such a modified class of matrices, we did not attempt to adapt the proof to the higher-order setup.
Consequently, while we observe the formation of metallic states at sharply terminated hinges for the discussed model of the layered RTP insulator, we cannot ensure the appearance of such metallic modes in other Hamiltonians with the same delicate Wannier invariant.

While we succeeded in extracting the anomalous hinge Berry phase of the layered RTP insulator by applying a suitable perturbation to the unit cells near the hinge, it is not clear from our analysis whether other choices of the hinge perturbation would yield the same result.
In particular, it is not a priori obvious whether a second hinge mode, if energetically detached by an additional Hamiltonian perturbation, would have a Berry phase of 0 or $\pi$.
If the answer can be $\pi$, then the net Berry phase of the two hinge modes would be trivial. Moreover, there is no reason to stop after the detachment of two hinge modes.
This makes the notion of the hinge Berry phase ambiguous, unless it is defined as the net Berry phase of a uniquely prescribed set of hinge-localized modes.
To resolve this prospective ambiguity, it is desirable to formulate a unique prescription which effectively classifies states into bulk-like, surface-like and hinge-like, by means of projected position operators for both $z$ and $y$ coordinates.
Our preliminary analysis, inspired by the approach to uniquely define the surface Chern number in Hopf insulators~\cite{Alexandradinata_prb}, suggests that such a generalization is indeed possible. 
However, since a self-contained discussion of the resulting bulk-to-surface-to-hinge correspondence requires introduction of novel physical concepts, we postpone its discussion to a separate work~\cite{Bzdusek2025}.

Further open questions relate to the concurrent-gap condition.
While this conjecture is observed to apply in sufficiently simple models~\cite{Khalaf}, including the layered RTP insulator in Eq.~(\ref{eqn:layered-RTP-Hamiltonian}), it is known that this conjecture does not apply generally~\cite{Yang:2020}.
It is at present not known what are the necessary or sufficient conditions that the Hamiltonian must obey for the concurrent-gap condition to hold.
Quantifying the extent to which this conjecture can fail would help better understand the phase diagram of the models of DWIs; in particular, how much the phase boundaries specified by closing of the surface energy gaps can differ from the phase boundaries determined by closing the Wannier gaps at $\nu = 1/2$.
Based on our numerical data, we anticipate that both types of phase boundaries intersect at a point in the phase diagram characterized by closing of the \emph{bulk} energy gap, meaning that these two sets of transition lines are not entirely independent.
A more systematic understanding of the concurrent-gap condition would also shed more light on the bulk-boundary correspondence of DWIs.

In a similar spirit, it would be valuable to understand, for both DIs and DWIs, how the topological boundary states are trivialized if one deviates from the mutually disjoint condition, e.g., through a weak coupling of additional orbitals that violate the disjointness condition.
Any realistic system (whether a real material or a synthetic/metamaterial implementation) has highly excited orbitals corresponding to arbitrary symmetry representation.
These high-energy orbitals could couple only weekly to the low-energy physics, but they cannot go completely absent. 
Therefore, understanding the effect of such perturbations on the bulk-boundary correspondence of DIs and DWIs is particularly important in the context of realistic experiments.
The importance of these questions in the context of concrete models is further amplified by the recently discussed relevance of delicate topological invariants in the context of strong photovoltaic responses~\cite{Alexandradinata:2024,Zhu:2024b,Jankowski:2024}.

While the layering construction can elevate any delicate topological invariant from Bloch bands to Wannier bands, we did not tackle the reverse question: do \emph{all} DWIs have a representative obtained by layering lower-dimensional DIs? 
Here, we anticipate the answer to be affirmative: given that the delicate topological obstruction of Wannier bands has an analog in the context of Bloch bands (which we have adopted as the defining property of DWIs), the layering construction ensures the elevation of the topology from Bloch bands to Wannier~bands.
It is interesting to speculate whether Wannier bands may, in addition, also exhibit a delicate topological obstruction that has no analog in the realm of energy bands, as such models would correspond to DWIs without a layering realization.

Finally, while the layering construction of DWIs applies in any spatial dimension $d$, it is interesting to consider whether it could also be extended in other ways.
One option, motivated by the higher-order aspect, is to iterate the layering construction, i.e., to consider $(d+2)$-dimensional models obtained by coupling alternating layers of $(d+1)$-dimensional DWIs. 
We expect the nontrivial Wannier topology of these models to be elevated to their \emph{nested} Wilson bands.
Alternatively, it is interesting to speculate whether the layering construction could be suitably extended to multi-gap topological invariants, such as the generalized quaternion charges of $\mathcal{PT}$-symmetric 1D chains~\cite{Wu:2018}, the $N$-band Hopf invariant in time-reversal breaking insulators~\cite{Lapierre_prr}, and similar~\cite{Lim:2023,Bouhon:2020,Slager:2024,Davoyan:2024}.
In analogy with DIs, these topological invariants are not visible to symmetry indicators but they are revealed by homotopy theory under a suitable redefinition of the gap condition. 
In this sense, multi-gap TIs provide a particular extension of DIs, which has not been considered in the present work.

\section{Acknowledgments}

We thank Andrea Kouta Dagnino, Yifei Guan, Aleksandra Nelson, and Titus Neupert for valuable discussions.
This work and the discussed models were originally presented at the APS Summit under the designation `boundary-obstructed delicate topological insulators'~\cite{Bzdusek2025}.
Z.G.~and T.B.~were supported by the Starting Grant No.~211310 by the Swiss National Science Foundation.
The code used to generate the presented data is openly accessible in Ref.~\onlinecite{supp}.

\appendix

\section{\texorpdfstring{\\}{}Topological characterization of 1D chains with two bands}
\label{sec:1d-chain-two-bands}

We here present the characterization of 1D chains with two bands and with $\mathcal{PT}$ and $\mathcal{P}$ symmetry in terms of the available stable and delicate topological invariants.
We assume throughout that we allow for breaking of the chiral symmetry (which is accidentally present if the Bloch Hamiltonian contains no term proportional to the identity matrix) as this symmetry is lost when generalizing to the multi-band case in Appendix~\ref{sec:1d-chain-three-bands}.

First, from the perspective of symmetry indicators, we find that the product of the inversion eigenvalues of the occupied Bloch states at the inversion invariant momenta takes values $\xi_\textrm{val}(0) \xi_\textrm{val}(\pi)= \pm 1$ [Eq.~(\ref{eqn:sym-ind}) on page~\pageref{eqn:sym-ind}]. 
The value $+1$ corresponds to a vanishing Berry phase ($\phi_\textrm{B} = 0$) along the 1D BZ, while the value $-1$ implies a nontrivial Berry phase ($\phi_\textrm{B} = \pi$). 
This symmetry-indicated $\mathbb{Z}_2$ invariant is \textit{stable} in the sense that it cannot be trivialized upon extending the occupied or unoccupied sector with bands that carry zero Berry phase.

Curiously, more topological information about the two-band chain can be obtained using homotopy theory~\cite{Bzdusek:2017,Brouwer}.
Due to the reality of the chain Hamiltonian, imposed by the space-time inversion $\mathcal{PT}$, the eigenstates can be expressed as real two-component vectors.
The eigenstates at each $k_x$ can be arranged as columns into an orthogonal $\mathsf{O}(2)$ matrix, which uniquely encodes the spectrally flattened Hamiltonian matrix [Eq.~(\ref{eqn:flattened-Hamiltonian}) on page~\pageref{eqn:flattened-Hamiltonian}].
However, this encoding is redundant; namely, each eigenstate exhibits an $\mathsf{O}(1)$ gauge degree of freedom, corresponding to an overall sign flip. 
Therefore the classifying space of two-band gapped real Hamiltonians is 
\begin{equation} 
\label{eqn:class-space-S1}
M_{(1,1)} = \mathsf{O}(2) / \mathsf{O}(1) \times \mathsf{O}(1) \simeq {S}^1.
\end{equation}
The latter expression can be obtained by considering $S^1\simeq \mathbb{R}P^1$ as a homogeneous space with the action of $\mathsf{O}(2)$, and it is naturally interpreted in terms of the spectrally normalized Hamiltonians expanded into the two available Pauli matrices.
It follows that the studied class of 1D Bloch Hamiltonians correspond to maps $\mathsf{T}^1 \mapsto M_{(1,1)}$, where $\mathsf{T}^1 \simeq S^1$ is the 1D BZ and $M_{(1,1)} \simeq S^1$ is the classifying space in Eq.~(\ref{eqn:class-space-S1}). 
Every such Hamiltonian corresponds to a unique element of the first homotopy group, $\omega \in \pi_1(M) = \mathbb{Z}$ [Eq.~(\ref{eqn:unstable-homotopy}) on page~\pageref{eqn:unstable-homotopy}], which is encoded by the winding number of the vector $(h_x,h_z)$ of the Hamiltonian $\mathcal{H}=h_x \sigma_x + h_z \sigma_z$ as the momentum $k_x$ is varied over the 1D BZ [Eq.~(\ref{eqn:winding-number-def}) on page~\pageref{eqn:winding-number-def}].
Within the two-band context, the inversion symmetry $\mathcal{P}$ ensures that the winding accumulated over the half-BZ $[0,\pi]$ is doubled over the complementary half-BZ $[\pi,2\pi]$.
If the symmetry indicator $\xi_\textrm{val}(0)\xi_\textrm{val}(\pi)$ is positive (negative), the winding number $\omega$ over the 1D BZ is ensured to be even (odd).

We find that homotopy theory reveals richer topological information than the symmetry indicator. 
This is expected as a matter of principle, because the homotopy group characterizes the Hamiltonian at all momenta inside the 1D BZ, whereas the symmetry indicator only looks at the two high-symmetry momenta.
However, the dichotomy between the two approaches is specific to two-band models: once the number of occupied ($n$) or unoccupied ($\ell$) bands grows to two or more, the first homotopy group of the corresponding classifying space 
\begin{equation}
\label{eqn:multi-band-classifying-space}
    M_{(n,\ell)} = \mathsf{O}(n+\ell)/\mathsf{O}(n)\times \mathsf{O}(\ell)
\end{equation}
reduces to $\mathbb{Z}_2$~\cite{Bzdusek:2017}, constituting the stable one-dimensional topological invariant of real vector bundles~\cite{Hatcher:2003}.
In this case, the winding number across the 1D BZ exactly reproduces the stable $\mathbb{Z}_2$ classification by the symmetry indicator $\prod_{j=1}^n \xi_j(0)\xi_j(\pi)$ (where $j$ labels the occupied bands), resp.~by the quantized Berry phase $\phi_{\textrm{B}}\in\{0,\pi\}$).
For $n>1$, $e^{i\phi_{\textrm{B}}}$ should be understood as the matrix determinant of the $n\times n$ Wilson loop~\cite{Alexandradinata:2014}.
In mathematical terminology, the $\mathbb{Z}$-valued homotopy group of $M_{(1,1)}$ is unstable. 
It follows that the two-band 1D chains with winding number $\omega \in 2\mathbb{Z}\backslash \{0\} = \{\pm 2, \pm 4, \ldots \}$ are delicate topological~\cite{Nelson_prl,Brouwer}.

\section{\texorpdfstring{\\}{}Topological characterization of 1D chains with many bands}\label{sec:1d-chain-three-bands}

When the total number of energy bands in the 1D chain is larger than two, the classifying space of $\mathcal{PT}$-symmetric insulating Hamiltonians is given by Eq.~(\ref{eqn:multi-band-classifying-space}). 
As a consequence, the topological characterization of multi-band 1D chains in terms of the homotopy group  $\pi_1[M_{(n,\ell)}]=\mathbb{Z}_2$ reproduces the classification in terms of the stable symmetry indicator.
However this $\mathbb{Z}_2$ classification is incomplete, because of the possibility of a nontrivial map from the half Brillouin zone $k_x \in [0,\pi]$ to $M_{(n,\ell)}$, conditioned by a proper arrangement of $\mathcal{P}$ eigenvalues at the high-symmetry momenta $k_x\in\{0,\pi\}$. Such maps are classified by the \emph{relative} homotopy groups~\cite{Hatcher:2002,Sun:2018}.

Our discussion of topological invariants in multi-band 1D chains is structured as follows.
First, we analyze in Appendix~\ref{sec:odd-winding-case} the effect of including a third band to a two-band model with an \emph{odd} winding number $\omega$, finding the absence of relative homotopy invariants.
We continue in Appendix~\ref{eqn:many-band-start-even} by discussing the addition of a third band to a two-band model with an \emph{even} winding number $\omega$.
We find that when the valence and conduction bands obey a mutually disjoint condition of inversion eigenvalues at $k_x \in \{0,\pi\}$, a refined delicate topological $\mathbb{Z}_2$ invariant can be computed over half-BZ. 
We argue that this $\mathbb{Z}_2$ invariant remains well-defined in models with arbitrarily many bands as long as the mutually-disjoint condition of the inversion eigenvalues is obeyed.

\subsection{Initial two-band model with odd winding number}
\label{sec:odd-winding-case}

We first consider two-band models whose occupied band obeys $\xi_\textrm{val}(0)\xi_\textrm{val}(\pi)=-1$, corresponding to nontrivial Berry phase $\phi_\textrm{B}=\pi$ (or, equivalently, to an odd winding number $\omega$).
For concreteness, we assume that the occupied Bloch eigenstate 
$\ket{u_\textrm{val}(k_x)}$ is $s$-like at $k_x = 0$ and $p$-like at $k_x=\pi$ (the case with flipped eigenvalues of $\mathcal{P}$ can be treated analogously).
The topology of the $\mathcal{PT}$-symmetric Hamiltonian is captured by the evolution of the occupied Bloch eigenstate as a function of $k_x$. 
This evolution can be visualized by plotting the \textit{real-valued} components $\braket{s}{u_\textrm{val}}$ and $\braket{p}{u_\textrm{val}}$ as a function of $k_x$, where $\ket{s}$ ($\ket{p}$) is the $s$-like ($p$-like) basis orbital of the two-band Hamiltonian. The reality of both inner products is a permissible gauge choice because the $\mathcal{PT}$-symmetric Hamiltonian is real at each $k_x$.
Due to the normalization of the real-valued $\ket{u_\textrm{val}(k_x)}$, the data point 
\begin{equation} 
(\braket{1}{u_\textrm{val}},\braket{2}{u_\textrm{val}}) \equiv \ket{u_\textrm{val}}
\end{equation}
lies on the unit circle for each $k_x$.
Owing to the $\mathsf{O}(1)$ gauge degree of freedom, antipodal points on the circle encode the same state, i.e., $\ket{u_\textrm{val}} = -\ket{u_\textrm{val}}$.
In addition, the inversion symmetry implies that complete information about the Bloch Hamiltonian can be obtained from the momentum range $k_x \in [0,\pi]$ (the half-BZ); therefore, we reduce the plots to this~range.

\begin{figure}
    \centering
    \includegraphics[width=0.99\linewidth]{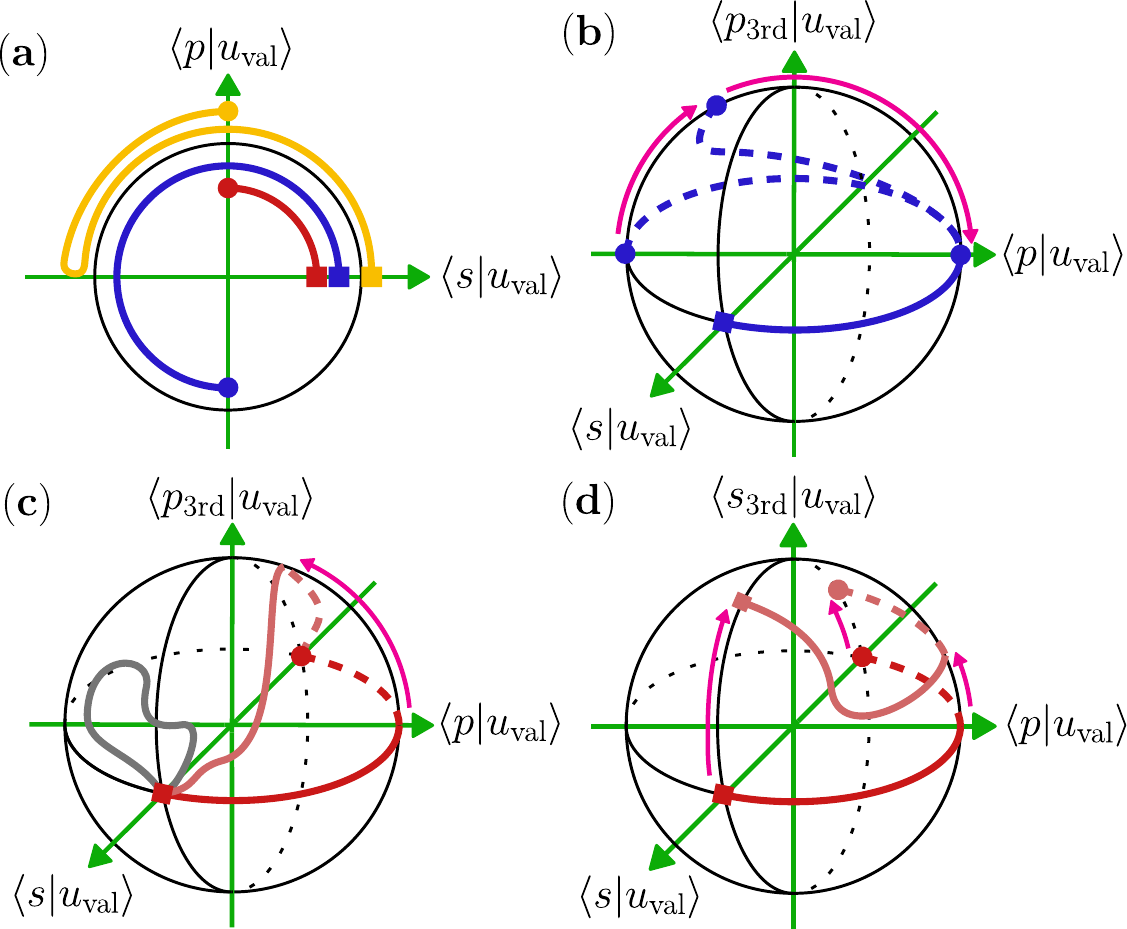}
    \caption{
    Paths representing the occupied Bloch state $\ket{u_\textrm{val}(k_x)}$ of inversion-symmetric 1D chains inside the half-BZ with range $k_x\in [0,\pi]$. 
    (\textbf{a})~For a two-band model with a band inversion (equivalently, with Berry phase $\phi_\textrm{B}=\pi$), the winding number $\omega$ across the 1D BZ is odd.
    The red and yellow paths subtend an oriented angle $\theta_\textrm{half-BZ}=\pi/2$ and represent winding number $\omega=1$ across the 1D BZ, while the blue path subtends $\theta_\textrm{half-BZ}=3\pi/2$ and represents $\omega = 3$. 
    (\textbf{b})~Extending such a two-band model by a third orbital facilitates a reduction of any odd winding number to $\omega=1$. 
    For example, the blue curve with $\omega=3$, shown in panel (\textbf{a}), is continuously deformed (magenta arrow) to a curve with $\omega=1$ upon the extension of the Hilbert space with an unoccupied $p$-like orbital.
    (\textbf{c})~For a two-band model without band inversion (equivalently, with Berry phase $\phi_\textrm{B}=0$), the winding number is even, $\omega \in 2 \mathbb{Z}$, allowing to define an integer-valued winding number $\omega_\textrm{half-BZ}\in \mathbb{Z}$ within the half-BZ $k_x \in [0,\pi]$.
    Extending such a two-band model while maintaining the mutually disjoint condition (here via the inclusion of an unoccupied $p$-like orbital) preserves the parity of $\omega_\textrm{half-BZ}$, allowing to introduce a delicate topological $\mathbb{Z}_2$ invariant protected by inversion symmetry. 
    A path representing the trivial (nontrivial) class is shown with grey (red) color. 
    (\textbf{d})~Violating the mutually disjoint condition (here via the inclusion of an unoccupied $s$-like orbital) results in a trivial classification for all paths with even winding~number~$\omega$. 
    }
\label{fig:homotopy_classes}
\end{figure}

For the specified two-band model, whose occupied eigenstate is $s$-like ($p$-like) at $k_x=0$ (at $k_x=\pi$), we have 
\begin{equation}   
\ket{u_\textrm{val}(0)} = \pm(1,0)\quad \textrm{and} \quad \ket{u_\textrm{val}}(\pi)= \pm(0,1).   
\end{equation}
In Fig.~\ref{fig:homotopy_classes}(\textbf{a}), we display several curves representing the possible evolution of $\ket{u_\textrm{val}}$ in the range $k_x\in[0,\pi]$.
For convenience, we fix the gauge at $k_x = 0$ (square end-point of the trajectories) to $+(1,0)$; then, assuming a continuous gauge on the half-BZ we can arrive at either $\pm(0,1)$ at $k_x = \pi$ (disk end-point of the trajectories).
The red and yellow paths in Fig.~\ref{fig:homotopy_classes}(\textbf{a}) both subtend an oriented angle $\theta_\textrm{half-BZ} = \pi/2$ on the circle. 
Upon including the complementary half-BZ, this angle is doubled to $\theta_\textrm{BZ}=\pi$ over the full 1D BZ.
In addition, since antipodal points on the circle encode the same Hamiltonian, we find that the classifying space of gapped two-band Hamiltonians ($S^1/\mathbb{Z}_2 = \mathbb{R} P^1 \simeq S^1$) is traversed once by this trajectory.
Therefore, the subtended angle $\theta_\textrm{half-BZ}=\pi/2$ represents winding number $\omega=1$ over the 1D BZ.

In contrast, the blue path in Fig.~\ref{fig:homotopy_classes}(\textbf{a}) subtends oriented angle $\theta_\textrm{half-BZ}=3\pi/2$, which after analogous considerations represents winding number $\omega = 3$ around the classifying space.
In the same spirit, any odd value of the winding number $\omega$ can be obtained under the specified inversion eigenvalues of the two bands.
Every odd winding number constitutes its own topological class, which remains well-defined as long as (1)~the $\mathcal{P}$ and $\mathcal{PT}$ symmetries are preserved and (2)~the Hilbert space is not extended with additional degrees of freedom.

We next consider the addition of a third band to the two-band model, such that the unoccupied band bundle has rank two. 
For concreteness, this third band is $p$-like at both high-symmetry $k_x$, implying it has trivial Berry phase~\cite{Alexandradinata:2014}. Applying Zak's relation between Berry phase and Wannier centers~\cite{Zak:1989},  the Wannier center of the third band (which is also the Wyckoff position of the added $p$ orbital, labeled $\ket{p_\textrm{3rd}}$) is at the center of the unit cell, which is the same placement as for the original two orbitals.
The occupied Bloch eigenstate is encoded as a normalized $3$-component real vector,
\begin{equation}
\ket{u_\textrm{val}}=(\braket{s}{u_\textrm{val}},\braket{p}{u_\textrm{val}},\braket{p_\textrm{3rd}}{u_\textrm{val}})
\end{equation}
whose evolution over the half-BZ traces a curve on a sphere ${S}^2$.
As in the two-band case, the gauge freedom implies that antipodal points on the sphere are identified.
This reduction reproduces the known classifying space ($M_{(1,2)}=S^2/\mathbb{Z}_2 = \mathbb{R}P^2$) of gapped $3$-band Hamiltonians with a rank-one occupied~band.

Due to the $p$-like character of the additional (third) orbital, the $p$-like occupied state $\ket{u_\textrm{val}(\pi)}$ is not pinned to a single pair of antipodal points $\pm(0,1,0)=\pm\ket{p}$ on the $S^2$; rather, the $p$-like character is compatible with the formation of any linear combination of $(0,1,0)=\ket{p}$ and $(0,0,1)=\ket{p_\textrm{3rd}}$.
This implies that the $p$-like occupied state $\ket{u_\textrm{val}(\pi)}$ can move freely along a big circle $S^1 \subset S^2$ in the plane spanned by $\braket{p}{u_\textrm{val}}$ and $\braket{p_\textrm{3rd}}{u_\textrm{val}}$.
Simple drawing reveals that any two-band model with odd winding number $\omega = 2n+1$ (which corresponds to subtended angle $\theta_\textrm{half-BZ}=n\pi + \tfrac{\pi}{2}$ on the big circle inside the plane spanned by $\braket{s}{u_\textrm{val}}$ and $\braket{p}{u_\textrm{val}}$) can be, by employing the $\braket{p_\textrm{3rd}}{u_\textrm{val}}$ dimension provided by the third band, continuously deformed into $\omega = 1$ (subtended angle $\theta_\textrm{half-BZ}=\pi/2$).
We illustrate such a deformation for $\omega = 3$ in Fig.~\ref{fig:homotopy_classes}(\textbf{b}).

The same reduction of the topological classification from odd winding number $\omega = 2n+1$ to a single topological class characterized by the symmetry indicator $\xi_\textrm{val}(0)\xi_\textrm{val}(\pi)=-1$ applies to the case when the Hilbert space is extended by an unoccupied $s$-like, an occupied $s$-like, or an occupied $p$-like energy band (in each case we assume the placement of the additional orbital in the center of the unit cell).
The reason is that, in each of these cases, one of the end-points of the image 
$\ket{u_\textrm{val}}: [0,\pi] \to S^2$ can move freely along a big circle $S^1\subset S^2$ (namely, the circle spanned by $\ket{s}$ and $\ket{s_\textrm{3rd}}$ if we include an $s$-like orbital, resp.~the circle spanned by $\ket{p}$ and $\ket{p_\textrm{3rd}}$ if we include a $p$-like orbital), thus facilitating a deformation analogous to the one shown in Fig.~\ref{fig:homotopy_classes}(\textbf{b}).
As a result, we find that for many-band models originating from a two-band model with an odd winding number (and with all orbitals placed at the center of the unit cell) there is no delicate topology, and that all such models fall into the same stable topological phase characterized by a negative symmetry indicator (resp.~by the nontrivial Berry phase).

\subsection{Initial two-band model with even winding number}
\label{eqn:many-band-start-even}

We next turn our attention to two-band models whose occupied band obeys $\xi_\textrm{val}(0)\xi_\textrm{val}(\pi)=+1$, corresponding to vanishing Berry phase $\phi_\textrm{B}=0$ (or, equivalently, to an even winding number $\omega$).
Unlike the case of odd winding numbers, analyzed in Appendix~\ref{sec:odd-winding-case}, supplementing the presently considered two-band models with a third orbital (assumed to be placed in the center of the unit cell) results in \emph{three} topologically distinct classes of gapped models. 
Specifically, we find two classes of Hamiltonians when we impose mutually disjoint inversion eigenvalues of the occupied and unoccupied bands, and one class when such a condition is violated.

For concreteness, we assume throughout the remainder of the text that the occupied band is $s$-like at both $k_x \in\{ 0,\pi\}$. [The discussion for a $p$-like occupied band, such as for the model in Eq.~(\ref{eqn:1D-chain-double-wind}) for $\beta=2$ as considered in Sec.~\ref{sec:1D-chain-single}, proceeds along the same logic.]
In contrast to the scenario of Sec.~\ref{sec:odd-winding-case}, the assumed $s$-like behavior at \emph{both} $k_x=0$ and $k_x=\pi$ ensures $\ket{u_\textrm{val}(0)}=+(1,0)$ and $\ket{u_\textrm{val}(\pi)}=\pm(1,0)$.
A curve that connects $(1,0)$ and $\pm(1,0)$ subtends an angle $\theta_\textrm{half-BZ}=n \pi$.
Upon including the complementary half-BZ, this angle is doubled to $\theta_\textrm{BZ}=2n\pi$, which, owing to the 
$\ket{u_\textrm{val}}=-\ket{u_\textrm{val}}$ equivalence, corresponds to an even winding number $\omega = 2n$ around the classifying space of the Hamiltonians as $k_x$ is tuned over the 1D BZ.

Importantly, due to the $\ket{u_\textrm{val}} = -\ket{u_\textrm{val}}$ equivalence, the trajectory of $\ket{u_\textrm{val}}$ over half-BZ is closed.
Therefore, it is possible to define the integer-valued winding number $\omega_\textrm{half-BZ}$ accumulated over $k_x\in [0,\pi]$. 
Since the winding number is doubled over the full 1D BZ, we find
\begin{equation}
    \omega_\textrm{half-BZ} = \tfrac{1}{2}\omega \in \mathbb{Z}.
\end{equation} 
Our next goal is to investigate how $\omega_\textrm{half-BZ}$ reacts to the addition of a third orbital to the two-band model.

First, we consider the case when the third band preserves the mutually disjoint condition~\cite{Nelson_prb}. 
With this we mean that the $\mathcal{P}$ eigenvalue of all occupied bands at both $k_x\in\{0,\pi\}$ is opposite of the $\mathcal{P}$ eigenvalue of all the unoccupied bands.
Such a situation is achieved by including either an occupied $s$-like orbital or an unoccupied $p$-like orbital. 
In the following, we opt for the latter option (the other case is analyzed similarly; the adaptation requires one to investigate the trajectory of the sole \emph{unoccupied} band).

The path of evolution of the sole occupied Bloch eigenstate $\ket{u_\textrm{val}(k_x)}$ with $k_x\in[0,\pi]$ in the three-band model is again visualized as a curve in $S^2$. 
Due to the $s$-like character of $\ket{u_\textrm{val}}$ at the half-BZ boundary $k_x\in\{0,\pi\}$, the end-points of the curve are restricted to isolated points: $\ket{u_\textrm{val}(0)} = \pm (1,0,0) = \ket{u_\textrm{val}(\pi)}$. 
Such paths are classified by elements of the relative homotopy group~\cite{Hatcher:2002} $\pi_1(S^2,\mathbb{Z}_2)$, which is equivalent to the regular homotopy group $\pi_1(\mathbb{R} P^2) = \mathbb{Z}_2$.
In Fig.~\ref{fig:homotopy_classes}(\textbf{c}), we illustrate representatives of the trivial (gray curve) and nontrivial (red curve) classes of this $\mathbb{Z}_2$ classification. 
Although the red curve can be continuously adjusted and deformed (magenta arrow in Fig.~\ref{fig:homotopy_classes}(\textbf{c})), the fact that it connects two antipodal points on $S^2$ implies that it cannot be continuously shrunk to point (in contrast to the gray curve).

Since the spectrally flattened Hamiltonians at momenta $k_x\in\{0,\pi\}$ are equal to each other, so that we can treat the half-BZ as a closed loop, the identified $\mathbb{Z}_2$ invariant can be interpreted as quantized Berry phase $\phi_\textrm{half-B}$ accumulated over the half-BZ [Eq.~(\ref{eqn:Berry-phase-half-BZ}) on page~\pageref{eqn:Berry-phase-half-BZ}].
Two-band models with $\omega = 4n+2$ across the full 1D BZ correspond to odd $\omega_\textrm{half-BZ}=2n+1$, and they reduce in the three-band setup to the nontrivial class with $\phi_\textrm{half-B}=\pi$.
In contrast, two-band models with $\omega = 4n$ across the 1D BZ give even $\omega_\textrm{half-BZ}=2n$, which in the three-band context translates to the trivial value $\phi_\textrm{half-B}=0$. 
To understand why $0$ vs $\pi$, consider that the definition of half-BZ Berry phase in Eq.~(\ref{eqn:Berry-phase-half-BZ}) requires a  \textit{quasi-periodic gauge}, defined by 
$\ket{u_\textrm{val}(k_x)}$ being differentiable and with $\ket{u_\textrm{val}(0)}=\ket{u_\textrm{val}(\pi)}$.
In the trivial case, this quasi-periodic gauge is compatible with the real gauge assumed in our pictorial arguments, and the Berry connection $\mathcal{A}_\textrm{val}(k_x) = \Im\braket{u_\textrm{val}}{\partial_{k_x}u_\textrm{val}}$ manifestly vanishes. 
In the nontrivial case, because the end points lie on antipodes differing in the sign of the wave function, the quasi-periodic gauge is incompatible with the real gauge; this means that if one insists on $\ket{u_\textrm{val}(k_x)}$ being differentiable with $\ket{u_\textrm{val}(0)}=\ket{u_\textrm{val}(\pi)}$, then $\ket{u_\textrm{val}(k_x)}$ is necessarily complex-valued for some interval of $k_x$, allowing for $\Im\braket{u_\textrm{val}}{\partial_{k_x}u_\textrm{val}}$ to be nonzero and  $\phi_\textrm{half-B}=\pi$.

Let us emphasize that this $\mathbb{Z}_2$ characterization survives in models with arbitrarily many bands as long as the mutually disjoint condition is preserved. 
This is because the mutually disjoint condition ensures that the spectrally flattened Hamiltonians at $k_x = 0$ and $k_x = \pi$ are equal to each other, allowing us to characterize the Hamiltonian on half-BZ, $\mathcal{H}: [0,\pi] \to M_{(n,\ell)}$, by a quantized Berry phase $\pi_1(M_{(n,\ell)})=\mathbb{Z}_2$. 
This quantized Berry phase should more generally be understood as the matrix determinant of the Wilson loop over half the Brillouin zone.

Finally, we show that the topological classification is trivialized when the two-band model with even winding number $\omega$ across the 1D BZ is supplemented with an orbital that \emph{violates} the mutually disjoint condition.
Such a situation arises if we include an occupied $p$-like or an unoccupied $s$-like energy band. 
In the following discussion, we consider the addition of an unoccupied $s$-like energy band (the other scenario is analyzed similarly; one then needs to investigate the trajectory of the \emph{unoccupied} band $\ket{u_\textrm{con}(k_x)}$).
The topology of the Hamiltonian is again encoded by following the path of evolution of $\ket{u_\textrm{val}(k_x)}$ on ${S}^2$ for momenta in the range $k_x \in [0, \pi]$.

By assumption, the occupied Bloch eigenstate $\ket{u_\textrm{val}}$ is $s$-like at $k_x = 0$ and $\pi$. 
Due to the inclusion of the additional $s$-like orbital $\ket{s_\textrm{3rd}}$, the $s$-like character of $\ket{u_\textrm{val}}$ at the half-BZ end-points is compatible with any linear combination of $(1,0,0) = \ket{s}$ and $(0,0,1)=\ket{s_\textrm{3rd}}$.
This implies that the end-points of the image $\mathcal{H}:[0,\pi]\to S^2$ can move freely along the big circle $S^1 \subset S^2$ spanned by $\braket{s}{u_\textrm{val}}$ and $\braket{s_\textrm{3rd}}{u_\textrm{val}}$.
However, any such path can be continuously shrunk to a point, which mathematically corresponds to the trivial relative homotopy group $\pi_1(S^2,S^1)=\mathbb{0}$~\cite{Sun:2018}.
We illustrate the general idea behind such a trivialization in Fig.~\ref{fig:homotopy_classes}(\textbf{d}), where the end-points of the red curve are first translated along the $s$-like big circle to the north pole of $S^2$, which is followed by the contraction of the closed loop based at the north pole.
This shows that violating the mutually disjoint condition results in a trivial classification where all the considered models fall into the same topological class with $\phi_\textrm{B} = 0$.

\bibliography{ref}

\end{document}